\def\mycmd{2}   

\if\mycmd1  
\documentclass[12pt,draftclsnofoot,onecolumn]{./IEEEtran}  
\else       
\documentclass[10pt,journal,letterpaper]{./IEEEtran}            
\fi         

\usepackage{dsfont}
\usepackage{cite}
\usepackage{arydshln}
\usepackage{subfigure}
\usepackage{amsmath}
\usepackage{amssymb}
\usepackage{amsthm}
\usepackage{commath}
\usepackage{url}
\usepackage{xcolor}
\usepackage{authblk}
\usepackage{algorithm}
\usepackage{hhline}

\usepackage{multirow}
\usepackage{array, booktabs}
\usepackage{chngpage}
\usepackage{tabularx}

\usepackage{enumitem}

\providecommand{\norm}[1]{\lVert#1\rVert}

\if\mycmd1  
\usepackage{setspace}
\else       
\fi         

\ifCLASSINFOpdf
  \usepackage[pdftex]{graphicx}
  \graphicspath{{./figures/}}
  \DeclareGraphicsExtensions{.eps}
\else
  \usepackage[dvips]{graphicx}
  \graphicspath{{./figures/}}
  \DeclareGraphicsExtensions{.eps}
\fi

\newtheorem{mypro}{Proposition}

\begin{document}

\title{Performance Analysis of License Assisted Access LTE with Asymmetric Hidden Terminals}

\author{Harim Lee, Hyoil Kim,~\IEEEmembership{Senior Member,~IEEE}, Hyun Jong Yang,~\IEEEmembership{Member,~IEEE}, Jeong Tak Kim, and SeungKwon Baek
\thanks{Harim Lee, Hyoil Kim, Hyun Jong Yang, and Jeong Tak Kim are with the School of Electrical and Computer Engineering, Ulsan National Institute of Science and Technology (UNIST), Ulsan, 44919, Korea (e-mail: \{hrlee,hkim,hjyang,precedo\}@unist.ac.kr), and SeungKwon Baek is with 5G Giga Communications Research Lab., Electronics and Telecommunications Research Institute (ETRI), Daejeon, 34129, Korea (e-mail: skback@etri.re.kr).
Hyoil Kim is the corresponding author.}
}

\IEEEcompsoctitleabstractindextext{     
\begin{abstract}
License Assisted Access (LAA) LTE (LTE-LAA) is a new type of LTE that aggregates the licensed LTE bands with the unlicensed bands via carrier aggregation.
To operate in unlicensed bands, LTE-LAA adopts the listen-before-talk policy and designs its channel access mechanism similar to WLAN's DCF.
In this paper, we consider an LAA network consisting of an LTE-LAA eNB coexisting with a Wi-Fi STA, and capture the {\em asymmetric hidden terminal problem} where the eNB recognizes the STA while the opposite is not true, which is caused by the asymmetric CCA thresholds between them.
We model the network as a joint Markov chain (MC) consisting of two individual MCs, and derive its steady-state probabilities, throughput, and channel access delay along with other key metrics like transmit, busy, collision, and doubling probabilities.
Through extensive evaluation, we confirm that the proposed model well predicts the dynamics of the LAA network, and identify important design guidelines for fair coexistence between LTE-LAA and WLAN as follows.
First, LTE-LAA should design its contention window (CW) doubling policy by considering Wi-Fi's packet duration and subframe-dependent collision probabilities.
Second, there exists a tradeoff between throughput and channel access delay, according to which the CW doubling policy should be adapted.

\end{abstract}

}

\maketitle
\IEEEdisplaynotcompsoctitleabstractindextext
\IEEEpeerreviewmaketitle

\section{Introduction}
License Assisted Access (LAA) LTE (LTE-LAA) is a new type of LTE proposed by the 3GPP, aiming at improving mobile quality of service (QoS) by increasing the network capacity.
Capacity enhancement is key to mobile QoS since the mobile traffic demand is sharply increasing and expected to become 8.2 times larger in 2020 than what it was in 2015 \cite{Cisco16}.
To cope with the trend, the 3GPP has been improving its features related to QoS in its recent releases \cite{3GPPTS23.207}, and LTE-LAA tries to address this by combining the unlicensed UNII bands at 5 GHz with LTE's licensed bands via the carrier aggregation (CA) functionality.



LTE-LAA introduces new features to realize its operation in the unlicensed bands.
For coexistence with legacy WLAN, LTE-LAA adopts the Listen Before Talk (LBT) policy \cite{3GPPTR36.889} 
along with Clear Channel Assessment (CCA), and designs its MAC mechanism similar to WLAN's Distributed Coordination Function (DCF).
Through adoption of LBT and DCF, LTE-LAA promotes more compatible and fair channel access with WLAN.
To better protect Wi-Fi devices, LTE-LAA also implements Discontinuous Transmission (DTX) that limits LTE-LAA's transmission duration by the Maximum Channel Occupancy Time (MCOT), which can be up to 10 milliseconds.

Compared to the conventional LTE, LTE-LAA faces new challenges in QoS provisioning because of its DCF-like channel access and DTX mechanisms.
Specifically, LTE-LAA should consider the random nature of how it utilizes spectrum, e.g., its inter-frame interval is inherently random and hard to predict whereas the conventional LTE has consecutive frames.
As a result, LTE-LAA should experience intermittent transmissions thus leading to time-varying network capacity.

Moreover, the LAA network is also confronted with a new phenomenon called {\em asymmetric hidden terminals}, which is caused by the asymmetry in setting the CCA threshold between LTE-LAA and WLAN.
Specifically, Wi-Fi STAs perform energy detection for undecodable signals with the threshold of -62 dBm, while LTE-LAA has a potential to detect Wi-Fi frames \cite{80211}.
As a result, when a Wi-Fi STA observes received signal strength (RSS) of LTE-LAA's signal less than -62 dBm, the STA believes that the channel is unoccupied; LTE-LAA, however, can detect the STA's signal and thus defers channel access.
This means that LTE-LAA can be hidden from Wi-Fi while the inverse is not true, and thus LTE-LAA may experience more delay in channel access than the research community currently believes.

Therefore, it becomes critical to provide a new analytical model that can accurately predict the behavior of LTE-LAA coexisting with WLAN.
%
The classical methods developed for WLAN's hidden terminal problem \cite{Ekici08,Hung10,Tsertou08}, however, are not applicable to the LAA scenario since they only considered symmetric hidden nodes.
Although a number of works have recently proposed analytical models that consider asymmetric hidden nodes \cite{Jeon14,Bhorkar14,Mvulla15}, none of them established a mathematical framework considering the DCF-nature of LTE-LAA.


This paper tries to propose a novel analytical framework for the network dynamics of LTE-WLAN coexistence, by taking LTE-LAA's unique features into account and modeling various performance metrics like throughput and channel access delay.
The contribution of this work is three-fold.
First, we modeled the LAA system as a joint Markov chain (MC) consisting of two individual MCs of LTE-LAA and WLAN, and have shown how they depend on each other.
Second, we derived two performance metrics of LTE-LAA and WLAN, i.e., average throughput and channel access delay, through which achievable QoS in coexistence can be correctly understood.
We also derived various system parameters related to such metrics, including transmit probabilities, busy probabilities, and per-packet and per-subframe collision probabilities.
To the best of the authors' knowledge, this is the first attempt to analyze subframe-dependent collision probabilities.
Finally, we confirmed the efficacy of the model through an extensive evaluation, and 
identified important design guidelines to promote fair coexistence between LTE-LAA and WLAN, as summarized as follows.
\begin{itemize}
\item LTE-LAA needs to re-design its contention window (CW) doubling policy, which currently takes account of the collisions in the first subframe only.  We found that the fairness in channel sharing is severely affected by which subframe is chosen for the policy and how large Wi-Fi's packets are.
\item There exists a tradeoff between throughput and channel access delay. For throughput enhancement, the CW doubling policy should prefer to use the last eligible subframe; for reduced delay, however, the first subframe is most beneficial.
\end{itemize}

The rest of the paper is organized as follows.
Section~\ref{sec:model} introduces our system model, and
Section~\ref{sec:analysis} analyzes the proposed joint MC (JMC) model and derives the throughput and delay of LTE-LAA and WLAN.
Section~\ref{sec:evaluation} investigates the accuracy of the model and presents the impact of asymmetric hidden nodes, via numerical evaluation.
Section~\ref{sec:discussion} discusses additional issues to consider for the LAA network, and
Section~\ref{sec:conclusion} concludes the paper.

\section{Related Work}
Due to its resemblance to WLAN, modeling LTE-LAA is highly relevant to WLAN's analytical frameworks.
Since the seminal work by Bianchi regarding saturated throughput \cite{Bianchi00}, more general frameworks have been proposed such as
the delay analysis in \cite{Ziouva02} and
the nonsaturated throughput analysis in \cite{Malone07}.
However, these work assumed a fully connected network thus ignoring the impact of hidden nodes.
Regarding the hidden terminal problem,
Ekici et al. \cite{Ekici08} and Hung et al. \cite{Hung10} introduced a modified Bianchi model, whereas
Tsertou et al. \cite{Tsertou08} proposed a new analytic framework.
However, these work assumed homogeneous packet lengths, which is inappropriate for the LAA network since MCOT may differ from Wi-Fi's packet duration.
Moreover, they cannot model the {\em multiple-collision phenomenon} caused by a Wi-Fi STA transmitting multiple times during an MCOT.
%
In addition, their models can only deal with symmetric hidden nodes, i.e., neither node A nor node B can detect each other.

Recently,
a number of researchers have proposed analytic models for the performance analysis of LTE-LAA \cite{Song16,Chen15,Cano16,Zhang15,Yin15,Jeon14,Bhorkar14,Mvulla15}.
Song et al.\cite{Song16} and Chen et al. \cite{Chen15} proposed MC-based mathematical frameworks, where
\cite{Song16} assumed a fully connected network and a fixed contention window for LTE-LAA while
Chen et al. \cite{Chen15} did not adopt LTE-LAA's backoff procedure. 
There also exist non-MC analytic models \cite{Cano16,Zhang15,Yin15}, but their approaches are limited to a fully connected network.
Among the few that considered asymmetric hidden nodes,
Jeon et al. \cite{Jeon14} and Bhorkar et al. \cite{Bhorkar14} ignored the binary exponential backoff feature of LTE-LAA, and
Mvulla et al. \cite{Mvulla15} did not present 
how to analyze the problem mathematically and is confined to the coexistence between IEEE 802.11ax and the legacy WLAN.
While aforementioned work dealt only with throughput performance, Cano et al. \cite{Cano16} derived the delay performance as well, but for WLAN only.

\section{System Model}
\label{sec:model}
\begin{figure}[!t]
\centering
\includegraphics[width=1\columnwidth]{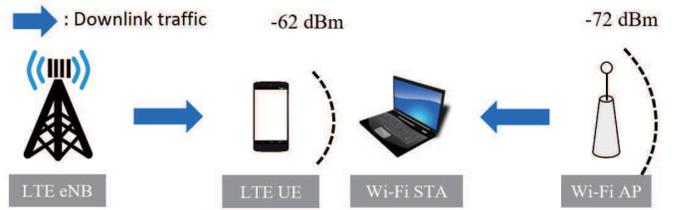}
\caption{LAA network model (eNB: node L, AP: node H)}
\label{Fig:System_Model}
\vspace{-0.1in}
\end{figure}

We consider an LAA network that consists of an LTE-LAA eNB (henceforth referred to as `node L') and a UE paired with the eNB, and
an WLAN that consists of a Wi-Fi AP (henceforth referred to as `node H') and a Wi-Fi STA paired with the AP.
Both networks are in the same unlicensed band.
%
We assume that there exists asymmetry between node L and node H such that
node L can detect node H's transmission while the opposite is not true.
Fig.~\ref{Fig:System_Model} illustrates the asymmetry, where
the dotted line marked with $-62$ dBm indicates the signal footprint where the received signal strength (RSS) of the eNB's transmission becomes $-62$ dBm, and
the dotted line marked with $-72$ dBm represents the coverage edge of the eNB.
Since node H is located beyond the $-62$ dBm line but within node L's coverage,
node H cannot detect node L's signal whereas node L can detect node H.

We assume that the LTE-LAA has downlink traffic only since the initial 3GPP standard considers such a scenario \cite{3GPP-RP-151569} and
so does the WLAN since the downlink traffic is expected to become 8 times heavier in 2020 than the uplink traffic \cite{ITU-Report15}.
Additionally, it is assumed that both of node L and node H have saturated traffic.

In our system model, we assume that if the eNB and the AP concurrently start their transmissions, they are in collision.
To show the validity of the assumption, we investigate the communicatable areas of the eNB and the AP provided that both are transmitting at the same time.
The communicatable area of the eNB (or the AP) is defined as the area within which the signal-to-interference-plus-noise ratio (SINR) of the UE (or the STA) is larger than the minimum SINR required for valid communications.
To derive such areas, we have utilized ITU InH pathloss model \cite{LTE-U-Forum} where
the minimum SINRs of the UE and the STA are set to 7 dB and 8 dB, respectively \cite{LTE_Small_Book}, and the transmit power of both nodes is set to 23 dBm \cite{ETSI301893}.
Under such conditions, in Fig. \ref{Fig:System_Model},
the distance between the eNB and the - 62 dBm line becomes 17 m while the distance between the eNB and the -72 dBm line becomes 37 m, meaning that the asymmetry occurs in the region (which will henceforth be referred to as the {\em asymmetry region}) covering from 17m to 37m.
%
%
\begin{figure}[!t]
\begin{center}
\subfigure[eNB's perspective]{
\includegraphics[width=0.466\columnwidth]{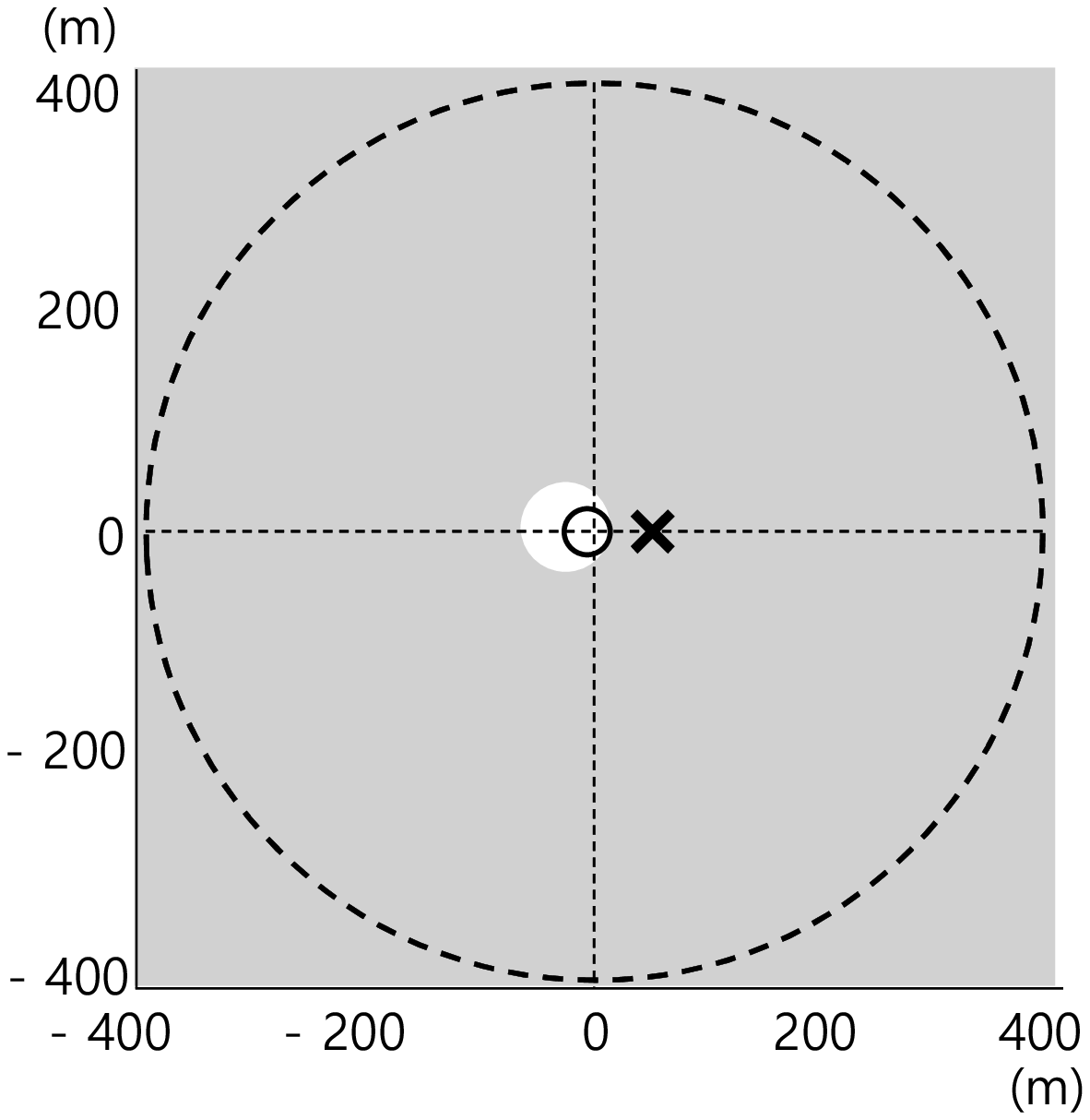}
\label{Fig:SINR_map_UE_w}
}
\subfigure[AP's perspective]{
\includegraphics[width=0.466\columnwidth]{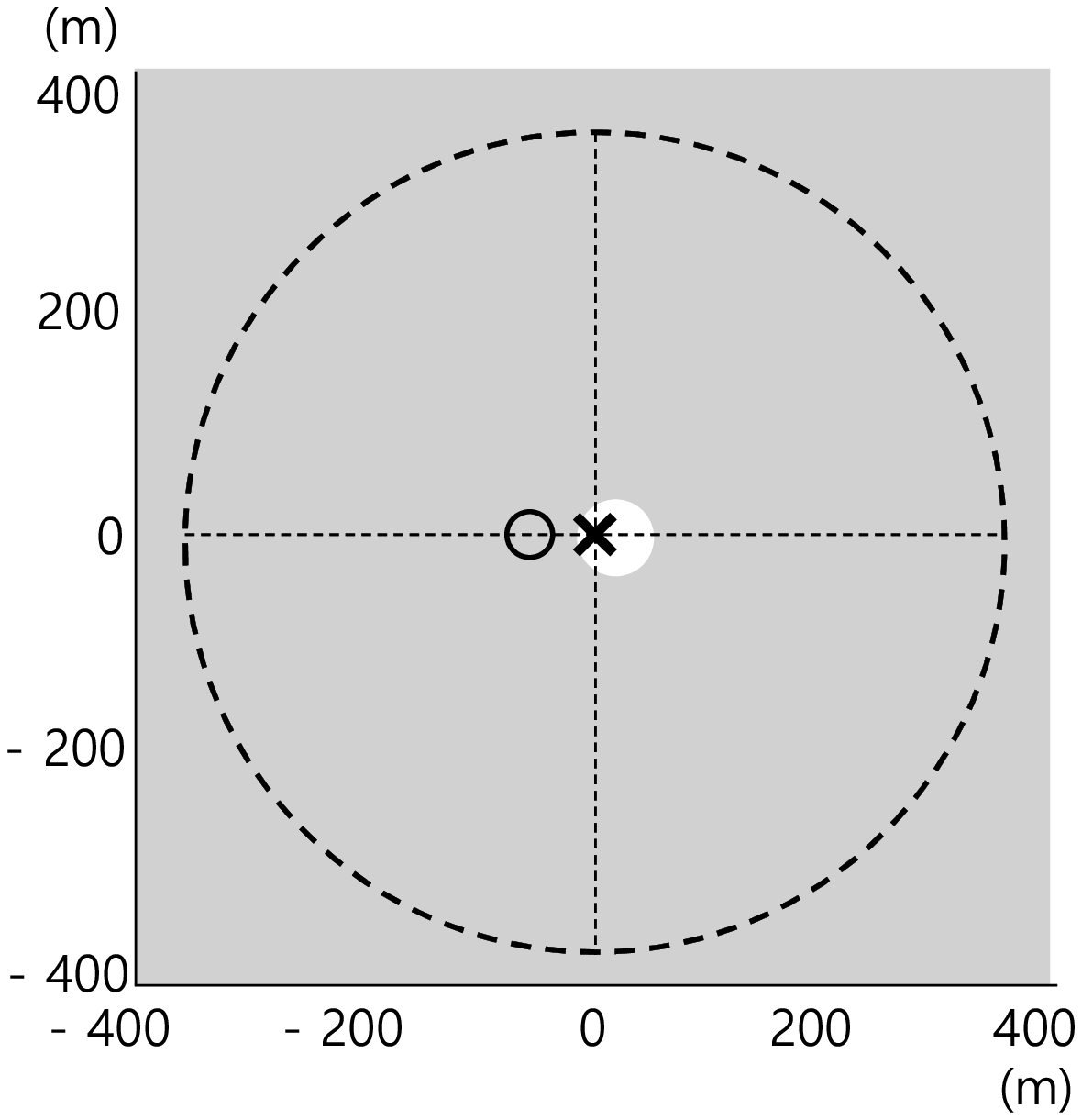}
\label{Fig:SINR_map_STA_w}
}
\caption{Communication possible area of eNB (circle) and AP (cross) when the distance between them is 20m}
\label{Fig:SINR_map}
\end{center}
\vspace{-0.14in}
\end{figure}

The Fig. \ref{Fig:SINR_map} shows the communicatable areas of the eNB and the AP in case the distance between them is 20 m, where the white regions are the communicatable areas of the eNB and the AP, respectively.
In addition, the dotted circle represents the coverage of the eNB (or the AP) assuming the AP (or the eNB) is not transmitting.
Since the communicatable areas are much smaller than the coverage areas, we can conclude that the eNB and the AP are almost always in collision whenever they transmit simultaneously.
%
\begin{table}[!t]
\centering
\caption{The ratio of two communication possible areas}
\label{Table:CPA}
\begin{tabular}{|c|c|c|}
\cline{1-3}
Distance & \multirow{2}{*}{eNB} & \multirow{2}{*}{AP} \\
between eNB and AP & & \\ \cline{1-3}
20m                                               & 0.92\%         & 0.79\%                            \\ \cline{1-3}
25m                                               & 1.43\%          & 1.24\%                 \\ \cline{1-3}
30m                                               & 2.07\%         & 1.79\%                      \\ \cline{1-3}
35m                                               & 2.88\%         & 2.43\%             \\ \cline{1-3}
\end{tabular}
\vspace{-0.07in}
\end{table}
Table~\ref{Table:CPA} shows that the same conclusion can be drawn for varying eNB-AP distances such as 20m, 25m, 30m, and 35m. The above discussion confirms the validity of our assumption.

Once node L starts transmission, it occupies the channel for one frame that is consisting of multiple subframes where each subframe lasts for 1 millisecond.
The number of subframes in a frame is determined by $T_{MCOT}$, the duration of the MCOT, and the subframes are indexed as $1, 2, \cdots$.
On the other hand, node H's packet duration is assumed to be geometrically distributed with $p_o = 1/T_\text{WiFi}$ being the probability that the transmission is complete in the current slot and $T_\text{WiFi}$ being the average packet length (in slots), which has been shown effective in modeling WLAN's traffic with mathematical tractability \cite{Babich10}.

\begin{table}[!t]
\centering
\caption{Channel access priority classes in LTE-LAA}
\label{Table:PriorityClass}
\begin{tabular}{|c|c|c|c|}
\cline{1-4}
Channel Access & \multirow{2}{*}{$CW_{min}$} & \multirow{2}{*}{$CW_{max}$} & \multirow{2}{*}{$T_{MCOT}$} \\
Priority Class & & & \\ \cline{1-4}
1                                               & 4          & 8          & 2 ms                         \\ \cline{1-4}
2                                               & 8          & 16         & 3 ms                         \\ \cline{1-4}
3                                               & 16         & 64         & 8 or 10 ms                  \\ \cline{1-4}
4                                               & 16         & 1024       & 8 or 10 ms                  \\ \cline{1-4}
\end{tabular}
\vspace{-0.07in}
\end{table}
The 3GPP has defined four priority classes in channel access \cite{3GPPTS36.213}, as shown in Table~\ref{Table:PriorityClass}.
Each class differently sets $CW_{min}$, $CW_{max}$, $T_{MCOT}$ where the first two parameters are related to LTE-LAA's DCF, and
a smaller-indexed class implies higher priority.
In classes 3 and 4, $T_{MCOT} = 10$ ms if the absence of WLAN can be guaranteed on a long term basis \cite{3GPPTS36.213}, and $T_{MCOT} = 8$ ms otherwise.
In this work, we assume that node L has priority class 4 traffic only, assuming low-priority data traffic.\footnote{The 3GPP has not yet discussed how to set the parameters for composite traffic, which is thus left as our future work.}
Although we should set $T_{MCOT} = 8$ ms due to the LTE-WLAN coexistence,
we consider two variations of $T_{MCOT}$, 8 ms and 10 ms, to show the impact of $T_{MCOT}$ on the coexistence performance and to suggest a more desirable value of $T_{MCOT}$.

Since LTE-LAA adopts DCF-like channel access, node L initially sets its CW as $CW_{min}$ and doubles it up to $CW_{max}$ at every collision.
In addition, the CW is initialized to $CW_{min}$ at successful transmission or when the maximum retry limit is reached.
There exists, however, a difference in how node L determines whether an MCOT period is in collision or successful.
In this regard, the 3GPP introduces a ratio-based rule (henceforth referred to as the {\em Ratio-rule}) in LTE Rel.13 \cite{3GPPTS36.213}, which is designed based on LTE's HARQ mechanism.\footnote{We assume that HARQ feedbacks are transmitted through the LTE licensed bands by the help of CA, since they are control packets.}
In the Ratio-rule, the ratio of the number of NACKs to the number of all HARQ feedbacks (i.e., ACKs and NACKs) is computed for the {\em reference subframe} (RSF) of an MCOT, and the CW is doubled at the completion of the MCOT if the thus-measured ratio is larger than the threshold.
Otherwise, the MCOT is regarded as successful.
In \cite{3GPPTS36.213}, the first subframe is chosen as the reference subframe.

In this work, we assume that if node H's transmission (at least partially) overlaps with the reference subframe of node L, the subframe is in collision and the CW doubles after the MCOT.
\begin{table}[!t]
\centering
\caption{Frequently-used Notations}
\label{Table:Notation}
\begin{tabular}{|m{1.5cm}|m{6.5cm}|m{0cm}}
\cline{1-2}
$L_t$, $H_t$	 						& The states of node L's and node H's MCs 	    			&\\[0.3ex] \cline{1-2}
$W_i$      		 						& Contention window size in backoff stage $i$   			&\\[0.3ex] \cline{1-2}
$b_{i,k}$, $b_{i,k^j}$						& Node L's stationary distributions in the backoff stage and the transmit stage, respectively &\\[0.5ex] \cline{1-2}
$p_b^L$, $p_d$	 						& Busy and doubling probabilities of node L 				&\\[0.7ex] \cline{1-2}
$p_b^H$, $p_c^H$						& Busy and collision probabilities of node H 				&\\[0.7ex] \cline{1-2}
$\tau_L$, $\tau_H$						& Transmit probabilities of node L and node H 				&\\[0.3ex] \cline{1-2}
$\tau_H^{\text{OW}}$, $\tau_H^{\text{MC}}$	& Node H's conditional transmit probabilities in an MCOT period and an OW period, respectively &\\[0.5ex] \cline{1-2}
$\pi_t$								& Stationary distribution of the simplified JMC 				&\\[0.3ex] \cline{1-2}
$\pi_t^l$								& Marginal distribution of the simplified JMC				&\\[0.7ex] \cline{1-2}	
$C_\text{sf}$							& Subframe collision probability 						&\\[0.3ex] \cline{1-2}
$S_L$, $S_H$							& Throughput of node L and node H 		&\\[0.3ex] \cline{1-2}
$D_L$, $D_H$						& Channel access delay of node L and node H 		&\\[0.3ex] \cline{1-2}
$T_{MCOT}$							& The duration of an MCOT period 						&\\[0.3ex] \cline{1-2}
$T_{\text{WiFi}}$								& The average packet length of node H 						&\\[0.3ex] \cline{1-2}
\end{tabular}
\vspace{-0.07in}
\end{table}
Table~\ref{Table:Notation} summarizes the notations used frequently throughout the paper.
This is reasonable since there exists a single UE in our model and thus all packets destined to the UE would be in collision when overlapped with node H's transmission.
In multi-UE scenarios, the assumption can be understood as a special case of the Ratio-rule where the threshold is set small enough so that a single NACK suffices to trigger the doubling condition.\footnote{Note that extension to the multi-UE scenario is our future work.}
Although the reference subframe should be the first subframe according to \cite{3GPPTS36.213}, we have set it as either the first or the last eligible subframe to show the impact of the chosen reference subframe.
Note that the last eligible subframe means the last subframe for which we can obtain the HARQ feedback considering the delay of 4 milliseconds (i.e., 4 subframes) to receive the feedback.
Therefore, depending on $T_{MCOT}$, we set the reference subframe as
\begin{itemize}
\item subframe 1 or 4, when $T_{MCOT} = 8$ ms, and
\item subframe 1 or 6, when $T_{MCOT} = 10$ ms.
\end{itemize}

We assume that time is slotted with the slot duration $T_{sl} = 9$ $\mu s$ as proposed by the 3GPP \cite{3GPPTS36.213}.
Note that $T_{sl} = 9$ $\mu s$ is commonly applied to LTE-LAA and WLAN.

\section{Performance Analysis}
\label{sec:analysis}
Let the state of the LAA system at slot $t$ be denoted by $Z_t = (L_t,H_t)$, where $L_t$ is the state of node L and $H_t$ is the state of node H.
Regarding its state transition, we make two assumptions as follows:

\begin{enumerate}[leftmargin=0cm,itemindent=.4cm,labelwidth=\itemindent,labelsep=0cm,align=left]
\item $\text{P}(L_{t+1}|L_t, H_t, ... , L_{0}, H_{0}) = \text{P}(L_{t+1}|L_t, H_t)$
\item $\text{P}(H_{t+1}|L_{t+1}, L_t, H_t, ... , L_{0}, H_{0}) = \text{P}(H_{t+1}|L_{t+1},L_t, H_t)$
\end{enumerate}

The first assumption is reasonable since node L determines its state transition according to the current channel state (i.e., busy or idle), which is determined by node H's behavior (i.e., transmitting or listening).\footnote{One exception is when $L_t$ is the last slot of an MCOT, where node L determines whether to double its CW or not according to the past collisions by node H within the MCOT. As a remedy to this, Section~\ref{subsection:doublingprobability} will introduce an average sense analysis by employing a constant doubling probability per MCOT.}
The second assumption implies that node H's state transition is affected by node L's state.
This is because node H's transmission can be collided with node L's concurrent transmission, even though node H does not see (and thus does not care) node L's behavior.

\begin{mypro}
\label{prop:MC}
Based on the assumptions 1 and 2, the LAA system becomes a Markov chain.
\end{mypro}
\begin{IEEEproof}
By the definition of conditional probability,
\begin{eqnarray}
\label{eqa:Proof_JMC_1}
\text{P}(Z_{t+1}|Z_t,\ldots,Z_{0}) &=& \text{P}(L_{t+1},H_{t+1}|L_t,H_t,\ldots,L_{0},H_{0}) \nonumber\\
&=& \text{P}(L_{t+1}|L_t, H_t, ... , L_{0}, H_{0}) \cdot \nonumber\\
&&  \text{P}(H_{t+1}|L_{t+1}, L_t, H_t, ... , L_{0}, H_{0}) \nonumber\\
&=& \text{P}(L_{t+1}|L_t,H_t) \cdot \nonumber\\
&&\text{P}(H_{t+1}|L_{t+1},L_t,H_t) \quad
\end{eqnarray}
where the last equality holds by the two assumptions.
By the law of total probability,
\begin{alignat}{1}
\label{eqa:Proof_JMC_2}
\text{P}(Z_{t+1}|Z_t) &= \text{P}(L_{t+1},H_{t+1}|L_t,H_t) \nonumber\\
&= \text{P}(L_{t+1}|L_t,H_t) \cdot \text{P}(H_{t+1}|L_{t+1},L_t,H_t)
\end{alignat}
where the last equality holds by the second assumption.
From Eqs.~(\ref{eqa:Proof_JMC_1}) and (\ref{eqa:Proof_JMC_2}), we obtain
\begin{alignat}{1}
\label{eqa:TransitionProbOfZt}
\text{P}(Z_{t+1}&|Z_{t},Z_{t-1},...,Z_{0})=\text{P}(Z_{t+1}|Z_{t})
\end{alignat}
which completes the proof.
\end{IEEEproof}

According to Proposition~\ref{prop:MC}, we model the LAA system as a joint MC (JMC) \cite{Bai11} of two individual MCs of node L and node H.
Specifically, we set $L_t$ as the outer MC with state space $\mathbb{S}_L$, whose transition affects the behavior of the inner MC $H_t$ with state space $\mathbb{S}_H$.
Then, the complexity in deriving the stationary distribution of $Z_t$ is proportional to the total number of states $|\mathbb{S}_L| \cdot |\mathbb{S}_H|$.

In the sequel, we first introduce node L's individual MC and show that $|\mathbb{S}_L|$ is quite large.
To reduce the complexity, we introduce a simplified MC of node L, and then analyze the JMC by combining the simplified MC with node H's MC.


\subsection{Markov Chain Analysis of Node L}
\label{subsec:analysis_LTE_MC}
\begin{figure}[!t]
\centering
\includegraphics[width=1\columnwidth]{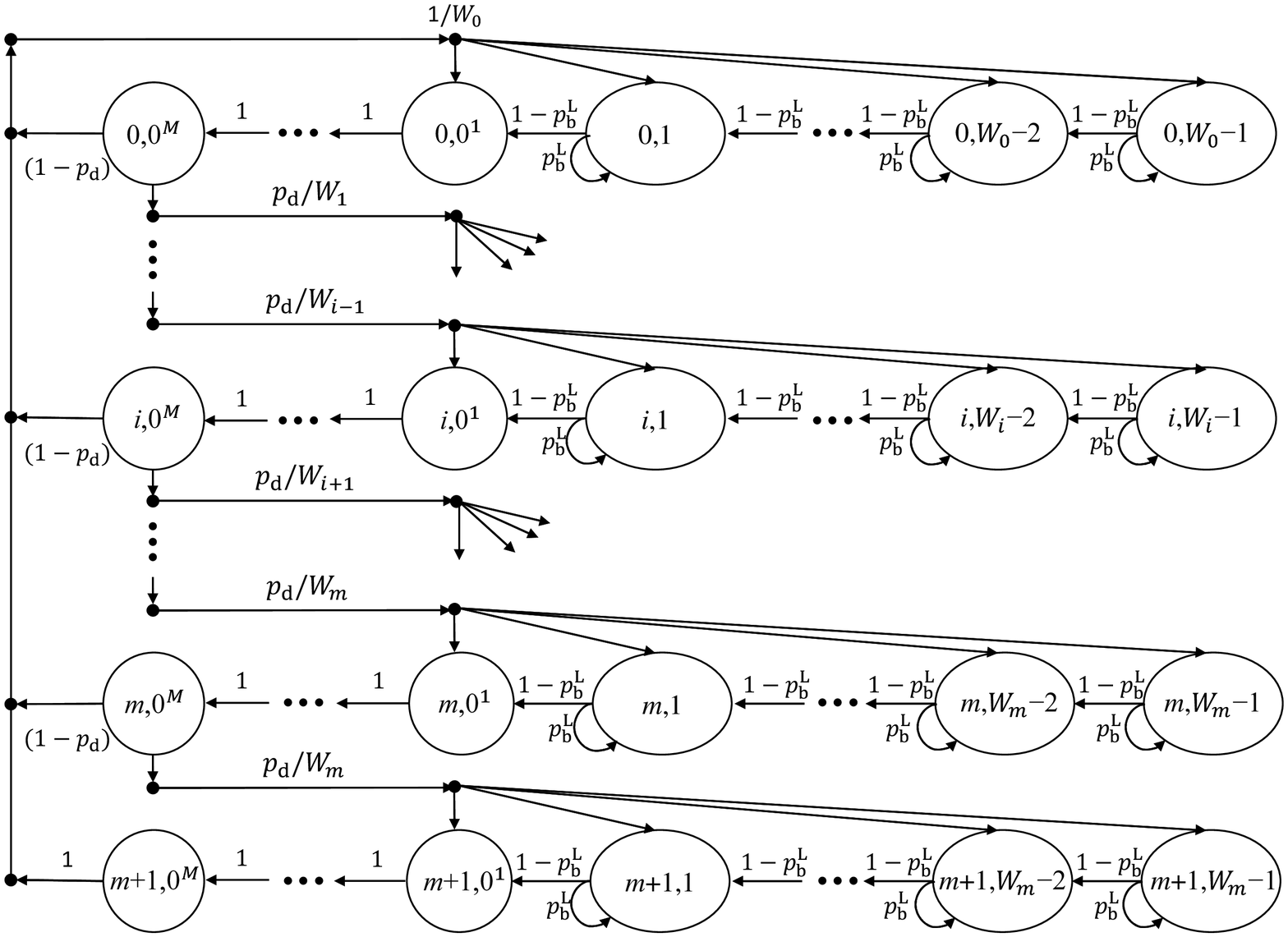}
\caption{Markov Chain of Node L (LTE-LAA eNB)}
\label{Fig:LTE-MC}
\end{figure}

The MC of node L is depicted in Fig.~\ref{Fig:LTE-MC}, whose state is denoted by either $(i,k)$ or $(i,0^j)$.
For $(i,k)$, $i$ is the backoff stage and $k$ is the backoff counter (BC), where $0 \le i \le m+1$ and $1 \le k \le W_i-1$ with $W_i = \min \{ 2^i, 2^m \} W_0$ and $W_0 = CW_{min}$.\footnote{Note that $W_m = W_{m+1}$.} 
Node L complies with DCF for channel access, while its transmission time is fixed as MCOT.
To capture this, the transmit stage consists of $M$ states where each state represents a physical slot of WLAN and $M$ is the length of MCOT in slots.\footnote{We have $M = T_{MCOT} / T_{sl}$, which is rounded off into an integer.}
Regarding this, state $(i,0^j)$ represents the $j$-th slot in the MCOT when node L is in stage $i$.
In the MC,
$p_b^L$ is the {\em average-sense} probability that node L senses the channel busy {\em per slot}, and
$p_d$ is the probability that node L doubles the CW after the completion of an MCOT (due to the collisions experienced during the MCOT).

If node L is in a backoff state, the BC is decremented by 1 whenever the channel is sensed idle with probability $(1-p_b^L)$, and stays the same otherwise.
That is,
\begin{alignat}{2}
P[(i,0^1)|(i,1)] & = 1-p_{b}^L,\quad k=1\nonumber\\
P[(i,k-1)|(i,k)] & = 1-p_{b}^{L},\quad k \in [2, W_{i}-1]\nonumber\\
P[(i,k)|(i,k)] & = p_{b}^{L},\quad k \in [1, W_{i}-1]\nonumber
\end{alignat}

The backoff stage is increased when doubling the CW is triggered, and reset to stage 0 otherwise.
In addition, it is reset to 0 in case the retry count limit is reached, i.e., collision occurs again at stage $(m+1)$.
Specifically, for $0 \leq i \leq m$ and $1 \leq k \leq W_{i+1}-1$,
\begin{eqnarray*}
P[(i+1,0^1)|(i,0^M)] &=& P[(i+1,k)|(i,0^M)] = \frac{p_d}{W_{i+1}} ,
\end{eqnarray*}
and for $0 \leq i \leq m$ and $1 \leq k \leq W_0-1$,
\begin{eqnarray*}
P[(0,0^{1})|(i,0^M)]   &=& P[(0,k)|(i,0^M)]   = \frac{1-p_d}{W_{0}}, \\
P[(0,0^{1})|(m+1,0^M)] &=& P[(0,k)|(m+1,0^M)] = \frac{1}{W_{0}} .
\end{eqnarray*}
%


Each transmit state lasts for a single slot and makes a transition to the next state with probability one.
The transition from state $(i,0^{j})$ can be derived as
\begin{eqnarray*}
P[(i,0^{j+1})|(i,0^{j})] & = & 1,\,\,\,\, i \in [0, m+1], \, j \in [1,M-1]
\end{eqnarray*}

Let $b_{i,k}$ and $b_{i,k^{j}}$ be the stationary distributions in the backoff state and in the transmit state, respectively.
Then,
\begin{eqnarray}
\label{eqnarray:Condition1}
b_{i,0^{1}} &=& \cdots = b_{i,0^{j}} = \cdots =  b_{i,0^{M}},\, i \in [0,m+1], \\
\label{eqnarray:Condition2}
b_{i-1,0^1} \cdot p_d &=& b_{i,0^M} \rightarrow
 b_{i,0^1} = (p_d)^i \cdot b_{0,0^1},\, i \in [1,m+1] .~~
 \phantom{1}
\end{eqnarray}
Owing to the chain regularities, $b_{i,k}$ can be derived as
\begin{equation}
\label{eqa:Final_b_i_k}
b_{i,k} = \frac{1}{1-p_{b}^{L}} \frac{W_i-k}{W_i} b_{i,0^{1}}, \quad 0 \le i \le m+1 .
\end{equation}

By Eqs.~(\ref{eqnarray:Condition2}) and (\ref{eqa:Final_b_i_k}), all stationary probabilities can be expressed in terms of $b_{0,0^{1}}$, $p_{b}^{L}$, and $p_d$.
Then, $b_{0,0^{1}}$ can be derived by using the normalization condition as follows:
\begin{equation}
\label{eqa:normalization}
1 = \sum_{i=0}^{m+1}\sum_{k=1}^{W_{i}-1}b_{i,k} + \sum_{i=0}^{m+1}\sum_{j=1}^{M}b_{i,0^j}
\end{equation}
By Eqs.~(\ref{eqnarray:Condition1}) and (\ref{eqnarray:Condition2}),
the two terms in the righthand side of Eq.~(\ref{eqa:normalization}) are rewritten as
\begin{alignat}{2}
\sum_{i=0}^{m+1} &\sum_{k=1}^{W_{i}-1}{b_{i,k}}
 = \sum_{i=0}^{m+1}\frac{(p_d)^i b_{0,0^1}}{1-p_b^L} \cdot \frac{W_i-1}{2} 
 = \frac{b_{0,0^1}}{2(1-p_b^L)} \cdot \nonumber\\
&\left[
 \frac{1-(2 p_d)^{m+1}}{1-2p_d} W_0
 - \frac{1-(p_d)^{m+1}}{1-p_d}
 + \frac{2^m W_0 - 1}{(p_d)^{-m-1}}
 \right] , \\
\sum_{i=0}^{m+1} &\sum_{j=1}^{M}b_{i,0^j}
 = \sum_{i=0}^{m+1} M \cdot (p_d)^i b_{0,0^1}
 = M \cdot b_{0,0^{1}}\frac{1-(p_d)^{m+2}}{1-p_d} .
\end{alignat}
%
%
As a result, $b_{0,0^{1}}$ is determined as
\begin{alignat}{2}
b_{0,0^{1}} &= \left[ \frac{1}{2(1-p_{b}^{L})} \left\{
 \frac{1-(2 p_d)^{m+1}}{1-2p_d} W_{0}
 - \frac{1-(p_d)^{m+1}}{1-p_d}
 \right. \right. \nonumber\\
&+ \left. (2^{m}W_{0}-1) (p_d)^{m+1} \Bigg\}
 + M \cdot \frac{1-(p_d)^{m+2}}{1-p_d} \right]^{-1} .
\end{alignat}

The {\em per slot} transmit probability of node L is denoted by $\tau_L$ and obtained as $\tau_{L} = \sum_{i=0}^{m+1}\sum_{j=1}^{M}b_{i,0^{j}}$, which can be expressed in terms of $b_{0,0^{1}}$ as
\vspace{-0.01in}
\begin{alignat}{1}
\label{alignat:tauL}
&\tau_{L} = M \cdot \frac{1-(p_d)^{m+2}}{1-p_d}
 b_{0,0^{1}}
\end{alignat}

\subsection{Markov Chain of Node H}
\begin{figure}[!t]
\centering
\includegraphics[width=1\columnwidth]{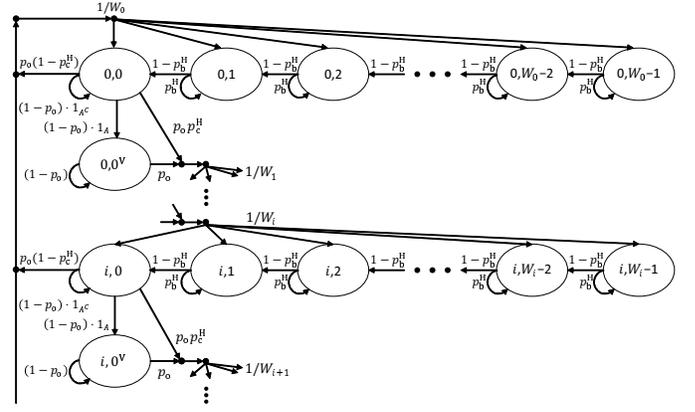}
\caption{Markov Chain of Node H (Wi-Fi AP)}
\label{Fig:MC_WiFi}
\end{figure}
In Fig.~\ref{Fig:MC_WiFi}, similar to node L, most of node H's states are denoted by $(i,k)$, where $i$ is the backoff stage and $k$ is the BC with $0 \le i \le m+1$, $0 \le k \le W_i-1$, and $W_i = \min \{ 2^i, 2^m \} W_0$.
In addition, $p_c^H$ denotes the {\em per-packet} collision probability, and $p_b^H$ denotes the {\em average-sense} probability that the channel is sensed busy {\em per-slot}, both from the viewpoint of node H.
\begin{figure}[!t]
\centering
\includegraphics[width=1\columnwidth]{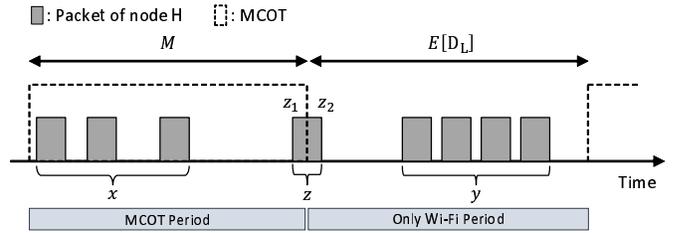}
\caption{MCOT period and Only Wi-Fi (OW) period}
\label{Fig:Collision_Probability}
\vspace{-0.1in}
\end{figure}

Node H's MC differs from node L's MC in the sense that it introduces new states denoted by $(i,0)^V, \forall i$.
To better describe the necessity of the new states, we define two types of periods, an MCOT period and an Only Wi-Fi (OW) period, as shown in Fig.~\ref{Fig:Collision_Probability}.
An MCOT period is the duration during which node L is in transmission, and an OW period is the duration during which node L is in the backoff stage and only node H can be in transmission.
Note that node H can transmit its packets in any of the two periods since it does not see node L's signal activities due to the asymmetry.

Since node H can transmit its packets in any of the two periods, a packet can be transmitted in an MCOT period.
The packet can stretch from the MCOT period to the following OW period (henceforth referred to as an overlapping packet)
such as the packet indicated by $z$ in Fig.~\ref{Fig:Collision_Probability}.
Because of the overlapping packets, there are two cases of node H's transmissions in an OW period such as
\begin{itemize}
\item Case 1: An overlapping transmission that starts {\em in an MCOT period} and completes {\em in the following OW period}.
\item Case 2: A transmission that starts {\em in an OW period} and completes {\em in the OW period}. 
\end{itemize}
In Case 1, the transmission overlaps an MCOT period and results in collision while the transmission of Case 2 is successful.
Therefore, collision probabilities of both cases are different.
To capture the different collision probabilities, node H's MC should distinguish the two cases of transmissions.
To represent the overlapping transmissions, the new states $(i,0)^V$ have been introduced to node H's MC.

For  state $(i,0)^V$, there are two incoming transitions and one outgoing transition.
Their transition probabilities are derived as follows.
One incoming transition from $(i,0)$ to $(i,0)^V$,$\forall i$ occurs when the period by node L switches from an MCOT period to the following OW period during a node H's transmission.
The switching is represented by using an indicator function as follows:
\begin{eqnarray}
1_A (L_t, L_{t+1}):= \begin{cases}
\label{eq:indicator}
1, & L_t \in S_{TX}^L,  L_{t+1} \in S_{BO}^L, \\
0, & \text{otherwise},
\end{cases}
\end{eqnarray}
where $S_{TX}^L$ is the set including all the transmission states in node L's MC and
$S_{BO}^L$ is the set including all the backoff states in node L's MC.
As a result, the transition probability from $(i,0)$ to $(i,0)^V$ is $(1-p_0) \cdot1_A(L_t, L_{t+1})$.
Another incoming transition occurs when the overlapping transmission is continuous with probability $1-p_o$.
The outgoing transition occurs when an overlapping transmission is complete with the probability $p_o$.

\subsection{Joint Markov Chain Analysis}
\begin{figure}[!t]
\centering
\includegraphics[width=1\columnwidth]{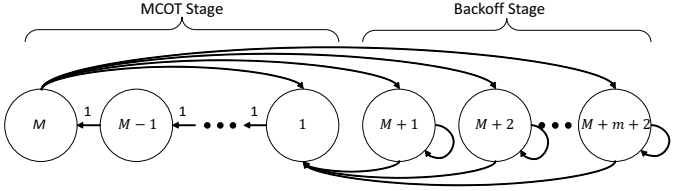}
\caption{Simplified Markov Chain of Node L}
\label{Fig:S_JMC}
\end{figure}
The joint MC is consisting of node L's MC in Fig.~\ref{Fig:LTE-MC} and node H's MC depicted in Fig.~\ref{Fig:MC_WiFi}.
With $T_\text{MCOT} = 8$ ms and $T_{sl} = 9$ $\mu s$,
the total number of transmit states in node L's MC becomes 7,040 and thus $|\mathbb{S}_L| >$ 10,000.\footnote{With $T_{MCOT} = 10$ ms, the total number of transmit states in node L's MC becomes 8,880.}
Therefore, it is desirable to reduce $|\mathbb{S}_L|$ to mitigate the complexity in analyzing the joint MC.
To do so, we introduce a simplified JMC (SJMC) that selectively groups the states of node L's MC in Fig.~\ref{Fig:LTE-MC} to produce the simplified MC shown in Fig.~\ref{Fig:S_JMC}.
Specifically, the transmit states $(i,0^j), \forall i$ in Fig.~\ref{Fig:LTE-MC} are combined into the single state $j$ in Fig.~\ref{Fig:S_JMC}, and
the backoff states $(i,k), \forall k$ in Fig.~\ref{Fig:LTE-MC} are combined into the single state $(i+M+1)$ in Fig.~\ref{Fig:S_JMC}.
Then, the stationary distribution of the simplified MC can be determined by using the stationary distribution of the original MC in Section~\ref{subsec:analysis_LTE_MC}.

Unless otherwise specified, 
$L_t$ will henceforth imply 
the state of node L's simplified MC at slot $t$.

\subsubsection{Transition Matrix of $\text{H}_{t}$ in SJMC}
\label{subsubsection:TM_Inner_MC}
\begin{figure}[!t]
\begin{center}
\subfigure[Transition matrix $P_H^{l,l'}$]{
\includegraphics[width=0.85\columnwidth]{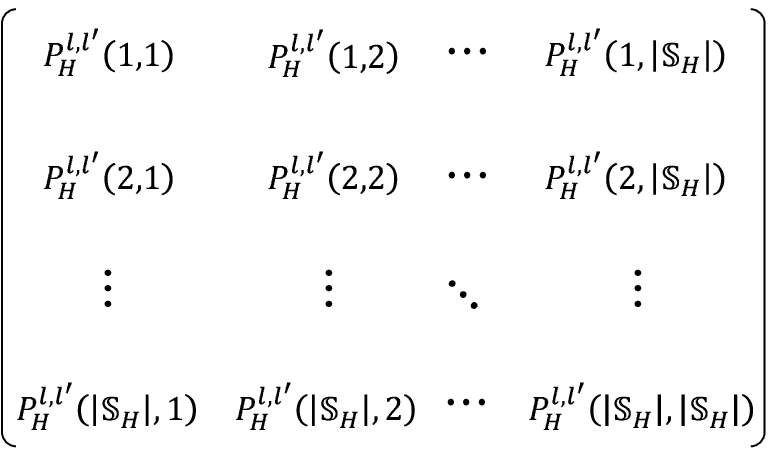}
\label{Fig:TM_Node_H}
}
\hspace{0.1in}
\subfigure[Diagonal matrix $D^{l,l'}$]{
\includegraphics[width=0.85\columnwidth]{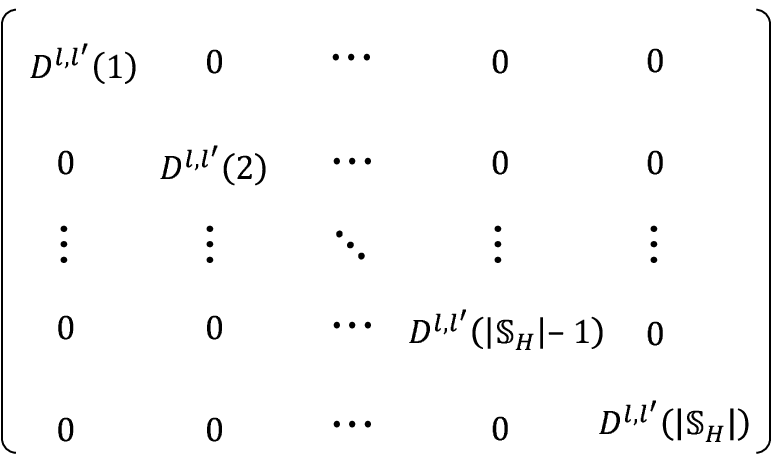}
\label{Fig:DiagMatrix}
}
\caption{Node H's transition matrix $P_H^{l}$ and diagonal matrix $D^{l,l'}$ (when $L_{t+1}=l'$ and $L_t=l$)}
\end{center}
\vspace{-0.1in}
\end{figure}
As introduced earlier, the JMC is consisting of the outer MC $L_t$ and the inner MC $H_t$.
The transition matrix $P_H^{l,l'}$ of the inner MC is consisting of the transition probabilities $P_H^{l,l'}(h,h') := P(H_{t+1}=h'|L_{t+1}=l', L_t=l,H_t=h)$ as illustrated in Fig.~\ref{Fig:TM_Node_H}.
$P_H^{l,l'}(h,h')$ is derived from the MC in Fig.~\ref{Fig:MC_WiFi}, where the two parameters $p_c^H$ and $p_b^H$ are set differently according to the state $L_t$ of the outer MC.

In the asymmetric coexistence, node H in the backoff stage decreases its BC at every slot since it does not recognize node L's signal, and a packet is transmitted once the BC hits zero.
Node H's transmission, however, has different consequences according to the type of periods it is transmitting in.
In an MCOT period, the transmission will always be in collision due to the ongoing transmission by node L.
The overlapping transmission that starts in an MCOT period is also in collision since it partially overlaps with an MCOT period.
On the other hand, any transmission initiating in an OW period will be successful, since node L can detect the signal and won't start a new MCOT period until the transmission completes.

Therefore, $p_c^H$ and $p_b^H$ are set as
$p_c^{H,\text{MC}} = 1$ and $p_b^{H,\text{MC}} = 0$ for an MCOT period, and $p_b^{H,\text{OW}} = 0$ and $p_c^{H,\text{OW}}=0$ for an OW period.
To capture the overlapping transmission, $1_A(L_t, L_{t+1})$ is set to one when $L_t = M$ and $L_{t+1} \in [M+1,M+m+2]$
while it is set to zero for the other transitions.
Thanks to the concept of the two periods, we only need to consider three transition matrices for the inner MC as follows:
\begin{itemize}
\item $P_{\text{OL}}$ with $ p_c^{H,\text{MC}}$, $p_b^{H,\text{MC}}$, and  $1_A(L_t, L_{t+1}) = 1$\\
when $L_t = M$, $L_{t+1} \in [M+1,M+m+2]$.
\item $P_{\text{OW}}$ with $ p_c^{H,\text{OW}}$, $p_b^{H,\text{OW}}$, and $1_A(L_t, L_{t+1}) = 0$\\
when $L_t \in [M+1,M+m+2]$, $L_{t+1} \in [1,M+m+2]$.
\item $P_{\text{MC}}$ with $ p_c^{H,\text{MC}}$, $p_b^{H,\text{MC}}$, and  $1_A(L_t, L_{t+1}) = 0$\\
for the other cases.
\end{itemize}
\subsubsection{Transition Probability of $\text{L}_{t}$ in SJMC}
\label{subsubsection:TP_Outer_MC}

According to Eq.~(\ref{eqa:Proof_JMC_2}),
the probability $P(L_{t+1}|L_t, H_t)$ is multiplied to the row of the inner MC's transition matrix in Fig~\ref{Fig:TM_Node_H} that corresponds to $H_t$.
This relationship can be expressed by exploiting a diagonal matrix $D^{l,l'}$ which consists of $D^{l,l'}(h) := P(L_{t+1}=l'|L_t=l,H_t=h)$, as shown in Fig.~\ref{Fig:DiagMatrix}.

The diagonal matrix is determined by considering the following five transition cases of the outer MC \footnote{In the other transitions not considered in the five transition cases, all the elements of $D^{l,l'}$ have zero since such transitions are impossible.}, such as
\begin{itemize}
\item Case 1: Transition {\em from} a state in the MCOT stage of Fig.~\ref{Fig:S_JMC} {\em to} the next state in the MCOT stage, i.e., $L_t \in \{ 1,\ldots,M-1 \}$ and $L_{t+1}=L_t + 1$.
\item Case 2: Transition {\em from} state $M$ {\em to} state 1, i.e., $L_t=M$ and $L_{t+1}=1$.
\item Case 3: Transition {\em from} state $M$ {\em to} a state in the backoff stage of Fig.~\ref{Fig:S_JMC}, i.e., $L_t = M$ and $L_{t+1} \in \{ M+1, \ldots, M+m+2 \}$.
\item Case 4: Transition {\em from} a state in the backoff stage {\em to} state 1, i.e., $L_t \in \{ M+1, \ldots, M+m+2 \}$ and $L_{t+1} = 1$.
\item Case 5: Transition {\em from} a state in the backoff stage {\em to} itself, i.e., $L_t = L_{t+1} \in \{ M+1, \ldots, M+m+2 \}$.
\end{itemize}
Case 1, 2, and 3 correspond to the situation where node L is in transmission, during which node L is not affected by the state of node H because node L keeps transmitting during an MCOT period once it has been started.
Therefore, all the elements of $D^{l,l'}$ have the same value regardless of $H_t$ such that $D^{l,l'}(h) = P_{l,l'}, \forall h$ where $P_{l,l'}$ is a constant for given $(l,l')$.
%
In Case 4 and 5, transitions are initiated when node L is in backoff states.
Since the operation of node L such as {\em defer} or {\em freeze} is affected by the state of node H, the elements of $D^{l,l'}$ should be derived according to $H_t$.

Then, $D^{l,l'}$ is determined as follows.

\noindent{\bf Case 1:}
In this case, $D^{l,l'}(h) = P_{l,l'}, \forall h$ and
$P_{l,l'} = 1$ since the transition occurs with probability 1 within an MCOT.%
\vspace{0.05in}

\noindent{\bf Case 2:}
In this case, $D^{l,l'}(h) = P_{l,l'}, \forall h$ and
$P_{l,l'}$ can be obtained by summing all the possible transitions from $(a,0^M)$ to $(b,0^1)$ where $a,b \in [0, m+1]$ in node L's MC.
Therefore,
\begin{equation}
P_{l,l'} = \left \{ \sum_{i=0}^{m} p_i^s (1-p_d) + p_{m+1}^s \right \} \frac{1}{W_0}
 + \sum_{i=0}^{m} p_{i}^s \cdot p_d \cdot \frac{1}{W_{i+1}},
\end{equation}
where $p_i^s = b_{i,0}^M/\sum_{a=0}^{m+1}b_{a,0}^M$ is the probability that node L is in backoff stage $i$ at the $M$-th MCOT slot.
%
\begin{figure*}[!t]
\begin{center}
\includegraphics[width=1.8\columnwidth]{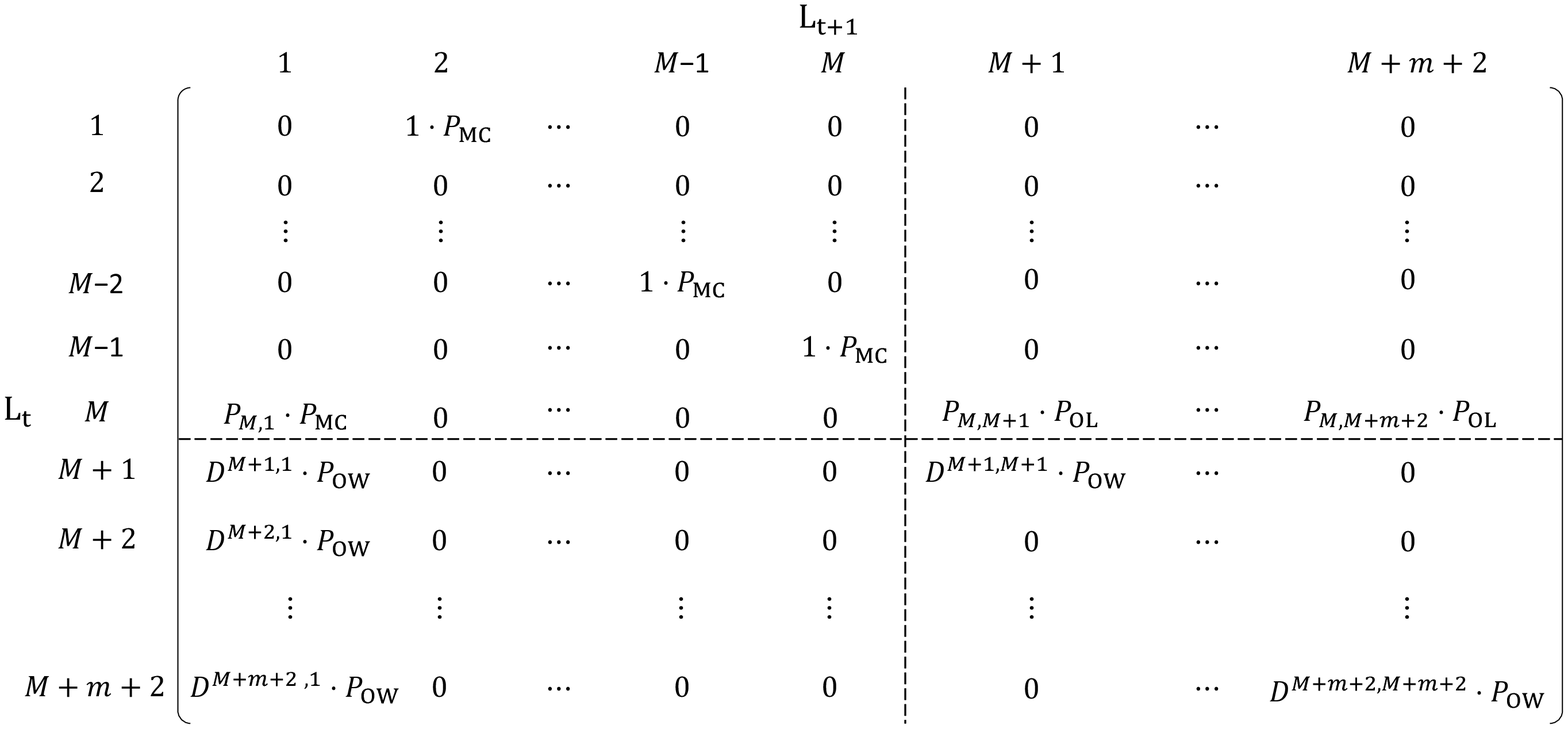}
\caption{Transition Matrix of SJMC}
\label{Fig:TM_SJMC}
\end{center}
\vspace{-0.1in}
\end{figure*}

\noindent{\bf Case 3:}
In this case, $D^{l,l'}(h) = P_{l,l'}, \forall h$.
When switching into state $M+1$ in the backoff stage, $P_{l,l'}$ is obtained by summing all the transitions from $(i,0^M)$ to $(0,k)$, $i \in [0,m+1]$, $k \in [1, W_0 - 1]$, in node L's MC.
When switching into state $M+1+i$, $i \in [1,m+1]$, in the backoff stage, $P_{l,l'}$ can be derived by considering all the transitions from $(a-1,0^M)$ to $(a,k)$, $a \in [1,m+1]$ $k \in[1, W_i - 1]$, in node L's MC.
Therefore,
\begin{eqnarray}
P_{l,l'} = \begin{cases}
\left \{ \sum_{i=0}^m  p_i^s (1-p_d) + p_{m+1}^s \right \}\frac{W_0 - 1}{W_0}, & \eta(l') = 0 , \\
p_{\eta(l'-1)}^s \cdot p_d \cdot \frac{W_{\eta(l')} - 1}{W_{\eta(l')}}, & \text{otherwise}, ~~
\end{cases}
\end{eqnarray}
where $\eta(l) = l - M - 1$, $M+1 \le l \le M+m+2$, is the backoff stage corresponding to state $l$ in node L's MC.
%

\noindent{\bf Case 4:}
In this case, $l$ is a state in the backoff stage and $l'=1$.
When node H is transmitting while node L is in the backoff stage, node L should freeze the backoff procedure, i.e., it stays at the backoff stage.
Therefore, we have $D^{l,l'}(h)=0$ when $h$ corresponds to a state of node H's transmit stage.
On the other hand, when both of node H and node L are in their backoff stage, node L starts transmission (i.e., it switches into state 1) once its BC becomes zero after decrement.
Therefore, $D^{l,l'}(h)$ with $h$ corresponding to node H's backoff stage
is the probability that the BC of node L happens to be one.\footnote{Please remind that state $i+M+1$, $0 \le i \le m+1$, of node L's simplified MC corresponds to the backoff stage $i$ of node L's original MC and its all possible BC values.}
As a result, $D^{l,l'}(h)$ is derived as
\begin{eqnarray}
D^{l,l'}(h) = \begin{cases}
\label{eq:case4}
0, & H_t \in S_{TX}^H, \\
\frac{b_{\eta(l),1}}{\sum_{k=1}^{W_{\eta(l)}- 1} b_{\eta(l),k}}, & \text{otherwise},
\end{cases}
\end{eqnarray}
where $S_{TX}^H$ is the set including all the transmission states in node H's MC.

\noindent{\bf Case 5:}
In this case, $l$ is a state in the backoff stage and $l'=l$.
When node H is transmitting while node L is in the backoff stage, node L should freeze the backoff procedure and thus $D^{l,l'}(h)=1$ where $h$ corresponds to node H's transmit stage.
When both of node H and node L are in their backoff stage, node L will stay in the same backoff state if the BC of node L does not become zero after decrement.
Hence, we have 
\begin{eqnarray}
D^{l,l'}(h) = \begin{cases}
\label{eq:case5}
1, & H_t \in S_{TX}^H, \\
1-\frac{b_{\eta(l),1}}{\sum_{k=1}^{W_{\eta(l)}- 1} b_{\eta(l),k}}, & \text{otherwise} .
\end{cases}
\end{eqnarray}

\subsubsection{Evolution of SJMC}
\label{subsubsec:SJMC}
By using the equations in Sections~\ref{subsubsection:TM_Inner_MC} and \ref{subsubsection:TP_Outer_MC}, the transition matrix $P^Z$ of the SJMC can be constructed as in Fig.~\ref{Fig:TM_SJMC}.
The evolution of the SJMC is given as $\pi_{t+1}=\pi_t \cdot P^Z$ where
\begin{eqnarray}
\label{eq:pi_t}
\pi_t &=& [\pi_t(1,1), \ldots, \pi_t(M+m+2,|\mathbb{S}_H|)], \\
\label{eq:pi_t_lh}
\pi_t(l,h) &=& P(L_t=l,H_t=h).
\end{eqnarray}
The marginal distribution $\pi_t^l$ is defined as $\pi_t^l=[\pi_t(l,1), \ldots, \pi_t(l,|\mathbb{S}_H|)]$.
Finally, the probability that node H is transmitting {\em per slot}, denoted by $\tau_H$, can be obtained as
\begin{equation}
\tau_H = \sum_{l=1}^{M+m+2}\pi_t^l\cdot m_H'
\end{equation}
where $m_H'$ is a column vector consisting of 0's and 1's that filters out node H's non-transmitting states. 
In addition, we can obtain the conditional probabilities, $\tau_H^{\text{MC}}$ and $\tau_H^{\text{OW}}$, that node H is transmitting {\em per slot} of an MCOT period and an OW period respectively, as follows:
\begin{equation}
\tau_H^{\text{MC}}= \frac{\sum_{l=1}^{M}\pi_t^l \cdot m_H'}{\tau_L}, \quad
\tau_H^{\text{OW}}= \frac{\sum_{l=M+1}^{M+m+2} \pi_t^l \cdot m_H'}{1 - \tau_L} .
\end{equation}
Furthermore, we notice $p_b^L = \tau_H^{\text{OW}}$ since $p_b^L$ equals the probability that node H is transmitting while node L is in its backoff stages.

$\tau_L$ was formerly derived as Eq.~(\ref{alignat:tauL}).
By using $\pi_t$, it can be also derived as
\begin{equation}
\tau_L =  \sum_{l=1}^{M} \norm{\pi_t^l}_1 ,
\end{equation}
where $\norm{\phantom{1}}_1$ is the 1-norm.


\subsubsection{Collision Probability of Node H}
\label{sec:NodeHcollision}

Fig.~\ref{Fig:Collision_Probability} describes a cycle consisting of an MCOT period and an OW period, where the MCOT period has a duration of $M$ and the OW period has a random duration $D_L$, both in slots.
On average,
there exist $x$ packets within an MCOT period and $y$ packets within an OW period, and there exist $z$ packets that stretch from an MCOT period to the following OW period. 
Note that $z < 1$ since there can be at most one overlapping packet per MCOT and not every MCOT will experience such a packet, and we denote the portion of $z$ in the MCOT period side by $z_1$ and that in the OW period side by $z_2$, where $z_1 + z_2 = 1$.

$p_c^H$ denotes the probability that a packet of node H is in collision.
Hence, the collision probability is measured per packet.
Using $x$, $y$, and $z$, we can express $p_c^H$ as
\begin{equation}
\label{equ:pcH_xyz}
p_c^H = \frac{x+z}{x+y+z} .
\end{equation}
In addition, by using $\tau_H$, $\tau_H^{\text{OW}}$, $\tau_H^{\text{MC}}$ and $\tau_L$, we have
\begin{equation}
\label{equ:tauHOW_xyz}
\frac{\tau_H^{\text{MC}} \cdot \tau_L}{\tau_H}  = \frac{x+z_1 \cdot z}{x+y+z}, \quad \frac{\tau_H^{\text{OW}}(1-\tau_L)}{\tau_H} = \frac{y+z_2 \cdot z}{x+y+z} .
\end{equation}
By Eqs.~(\ref{equ:pcH_xyz}) and	 (\ref{equ:tauHOW_xyz}), $p_c^H$ are derived as
\begin{eqnarray}
\label{eq:pcH}
p_c^H &=& \frac{x+z_1z+z_2z}{x+y+z}
 = \frac{\tau_H^\text{MC} \cdot \tau_L}{\tau_H}
 + z_2 \cdot p_\text{overlap} ,
\end{eqnarray}
where $x+y+z = (M + E[D_L]) \tau_H/T_\text{WiFi}$, and
$p_\text{overlap} := z / (x+y+z)$ is the probability that a packet of node H overlaps with both an MCOT period and the following OW period.

To derive $p_\text{overlap}$, we let
\begin{itemize}
\item $V$ denote the {\em overlap event} where the packet node H is transmitting overlaps with an MCOT period and the following OW period,
\item $\widetilde{T}_L$ denote an event that node L is transmitting for $T_{MCOT}$,
\item $\widetilde{T}_H$ denote an event that node H is transmitting a packet.
\end{itemize}
Then, $p_\text{overlap} = P[V | \widetilde{T}_H]$ by definition, from which we derive
\begin{eqnarray}
\label{eq:tauHcross}
p_\text{overlap} &=& P[V | \widetilde{T}_H]
 = \frac{P[V, \widetilde{T}_H]}{P[\widetilde{T}_H]} \\
&=& \frac{
 P[V, \widetilde{T}_H, \widetilde{T}_L]
 + P[V, \widetilde{T}_H, \widetilde{T}_L^{\mathsf{c}}]
 }{\tau_H}
 = \frac{P[V, \widetilde{T}_H, \widetilde{T}_L]}{\tau_H} \phantom{111}
 \nonumber
%
\end{eqnarray}
where the superscript $\mathsf{c}$ means a complement, and
the last equality holds since the overlap event cannot occur when node L is in the backoff stage, i.e., $P[V, \widetilde{T}_H | {\widetilde{T}_L}^{\mathsf{c}}] = 0$.

Since $\widetilde{T}_L$ includes $M$ possibilities, i.e., $L_t$ is one of $M$ transmit states, the numerator of Eq.~(\ref{eq:tauHcross}) is written as
\begin{alignat}{1}
\label{eqa:TotalProb2}
P[V, \widetilde{T}_H, \widetilde{T}_L] =&
 \sum_{i=1}^{M} \Big\{
 P[\widetilde{T}_L]
 \cdot P[L_t=i | \widetilde{T}_L]
 \cdot P[\widetilde{T}_H | \widetilde{T}_L, L_t=i] \nonumber\\
\cdot& P[V | \widetilde{T}_H, \widetilde{T}_L, L_t=i] \Big\} \nonumber\\
=& \frac{\tau_L \cdot \tau_H^\text{MC}}{M}
 \sum_{i=1}^{M} (1-p_o)^{M-i+1} \nonumber\\
\approx& \frac{\tau_L \cdot \tau_H^\text{MC}}{M}
 \sum_{i=1}^{\infty} (1-p_o)^{M-i+1} \nonumber\\
=& \frac{\tau_L \cdot \tau_H^\text{MC} \cdot (T_\text{WiFi}-1)}{M} .
%
\end{alignat}
In deriving Eq.~(\ref{eqa:TotalProb2}), the second equality holds because node H's packet duration follows geometric distribution and
\begin{alignat}{1}
\label{eqa:Prob_Node_L_TX}
P[\widetilde{T}_L] &= \tau_L , \\
\label{eqa:Prob_on_i_slot}
P[L_t=i | \widetilde{T}_L] &= 1/M , \\
\label{eqa:Prob_Node_H_TX_on_i_slot}
P[\widetilde{T}_H | \widetilde{T}_L, L_t=i] &= \tau_H^\text{MC} ,
\end{alignat}
where Eq.~(\ref{eqa:Prob_on_i_slot}) is from Eq.~(\ref{eqnarray:Condition1}).
In addition, the approximation of $M = \infty$ is made since $M \gg 1$.


As a result, $p_c^H$ is determined as
\begin{alignat}{1}
p_c^H &=
\frac{\tau_H^\text{MC} \cdot \tau_L}{\tau_H}
+ \frac{\tau_L \cdot \tau_H^\text{MC} \cdot (T_\text{WiFi}-1)}{2 \cdot \tau_H \cdot M} ,
\end{alignat}
where $z_2 = 1/2$ is assumed.
Note that the assumption of $z_1 = z_2 = 1/2$ is reasonable since
the completion of node H's transmission does not concern when an MCOT period ends,\footnote{Please remind that node H cannot see node L's signal activity.}
and node L completes its MCOT after a fixed amount of $T_{MCOT}$ regardless of whether there exists node H's ongoing transmission or not.
In other words, once the overlap event occurs, it can happen at any slot within node H's packet duration, i.e., uniformly.
In Section~\ref{subsec:cbdPr}, we will show that this assumption is proven to be valid in most cases.


\subsubsection{Doubling Probability of Node L}
\label{subsection:doublingprobability}
Doubling of node L's CW is triggered when the RSF is in collision, where the collision implies that at least one slot of the subframe is in collision due to concurrently transmitting node H.
%
Since node H always decrements its BC at every slot while not transmitting,
the doubling probability $p_d$ becomes the probability that node H's BC is smaller than the length of a subframe (in slots) at the start of the RSF, such as
\begin{alignat}{1}
p_d &= C_{\text{sf}}(r), \text{ when RSF = $r$}, \\
C_{\text{sf}}(r) &= \sum_{ \{ h | 1 \leq \mathcal{BC}(h) < \text{SF}_{slot} \} }
 \frac{ \pi_t^{(r-1) \cdot \text{SF}_{slot}+1} (h) }
      { \norm{ \pi_t^{ (r-1) \cdot \text{SF}_{slot}+1} }_1 } ,
\end{alignat}
where
$\mathcal{BC}(h)$ is the BC value corresponding to state $h$ of node H's MC,
$\pi_t^l$ is the marginal distribution as previously defined in Section~\ref{subsubsec:SJMC}, and
$\text{SF}_{slot}$ is the length of a subframe in slots.
%
Note that $C_{\text{sf}}(r)$ will be called the {\em subframe collision probability} for subframe $r$.

\subsection{Throughput and Access Delay Analysis}
\subsubsection{Node L}

Node L's normalized throughput $S_L$ is defined as the time proportion of successful transmissions, which is given as
\begin{equation}
\label{equation:ThroughputLTE}
S_L = \frac{\alpha \cdot M}{M + E[D_L]} = \alpha \cdot \tau_L, \quad
\alpha = \frac{ \sum_{r=1}^{n_{\text{sf}}} \text{C}_{\text{sf}}(r) }
              { n_{\text{sf}} } ,
\end{equation}
where $E[D_L]$ is node L's average channel access delay in slots,
$\alpha$ is the ratio of successful subframes, and 
$n_{\text{sf}}$ is the number of subframes in an MCOT. 

$E[D_L]$ is the expected time between two consecutive MCOTs, which is derived from Eq.~(\ref{equation:ThroughputLTE}) such as
\begin{equation}
E[D_L] = \frac{1 - \tau_L}{\tau_L} \cdot M.
\end{equation}

\subsubsection{Node H}

Node H's normalized throughput $S_H$ is defined as the time proportion of successful transmissions, which is given as
\begin{equation}
\label{equation:ThroughputWiFi}
S_H = \frac{ (1-p_c^H) \cdot T_{\text{WiFi}} }{ T_{\text{WiFi}} + E[D_H] }
    = (1-p_c^H) \cdot \tau_H ,
\end{equation}
where $E[D_H]$ is node H's average channel access delay in slots.

$E[D_H]$ is the expected time between two consecutive Wi-Fi transmissions, which is derived from Eq.~(\ref{equation:ThroughputWiFi}) such as
\begin{equation}
E[D_H]= \frac{1-\tau_H}{\tau_H} \cdot T_\text{WiFi}
\end{equation}

\section{Performance Evaluation}
\label{sec:evaluation}
We have verified the accuracy of the proposed model using our MATLAB-based DCF simulator that implements the key functional requirements of LTE-LAA and WLAN.
For node H, five per-packet transmit durations are considered such as 4, 54, 104, 154, and 204 slots,
where the minimum duration of 4 corresponds to a 350-byte long packet transmitted at a data rate of 78 Mbps and
the maximum duration of 204 corresponds to a 1,500-byte long packet transmitted at a data rate of 6.5 Mbps.
For node L, $T_{MCOT}$ is set as either 8 milliseconds or 10 milliseconds, and the RSF is set accordingly as follows:
\begin{itemize}
\item When $T_\text{MCOT} =$ 8 ms:  RSF is either 1 or 4,
\item When $T_\text{MCOT} =$ 10 ms: RSF is either 1 or 6.
\end{itemize}
For both nodes L and H, we set $CW_{min}=16$, $CW_{max}=1024$, and $m=6$, according to IEEE 802.11ac \cite{80211ac} and LTE Rel.13 \cite{3GPPTS36.213}.

To show the accuracy of the proposed model and the presented analysis, we measure each of the derived metrics by our simulator and compare it with its analytic result.
The metrics investigated include $\tau_L$, $\tau_H$, ${\tau}_H^{\text{OW}}$, ${\tau}_H^{\text{MC}}$, $p_c^H$, $p_b^L$, $p_d$, $C_{\text{sf}}$, $S_L$, $E[D_L]$, $S_H$, and $E[D_H]$.
Our analysis in the previous sections revealed that every metric is a function of $\pi_t$, 
where $\pi_t$ is again
a function of $p_b^L$ and $p_d$.
Therefore, we first measure $p_b^L$ and $p_d$ by the simulator, and then determine $\pi_t$ 
analytically by using Eq.~(\ref{eq:pi_t}).
Next, using thus-obtained $\pi_t$, 
we analytically determine each of the other metrics (including $p_b^L$ and $p_d$ as well) according to its corresponding equation derived earlier.

The aforementioned approach is taken since it helps us focus on evaluating the accuracy of the derived equations.
As an alternative way, all metrics can also be determined via an iterative numerical evaluation as follows.
Since $\pi_t$ is a function of $p_b^L$ and $p_d$ and each of $p_b^L$ and $p_d$ is a function of $\pi_t$,
we can notice that $\pi_t$ is a function of itself.
Therefore, the numerical analysis can be formed with the following iterative steps:
\begin{itemize}
\item {\bf Initialization step}: $\pi_t$ is obtained by using any seed values of $p_b^L$ and $p_d$.
\item {\bf Step 1}: $p_b^L$ and $p_d$ are calculated by using $\pi_t$ obtained in the previous step.
\item {\bf Step 2}: $\pi_t$ is calculated by using $p_b^L$ and $p_d$ obtained in the previous step.
    Go to Step 3 if $\pi_t$ reaches convergence, e.g., the gap between new $\pi_t$ and previously-calculated $\pi_t$ becomes less than a given threshold.
    Otherwise, go back to Step 1.
\item {\bf Step 3}: Calculate all other metrics using $\pi_t$.
\end{itemize}


\subsection{Transmit Probabilities}
\label{subsec:TxPr}
We consider four types of transmit probabilities,
$\tau_L$, $\tau_H$, $\tilde{\tau}_H^{\text{OW}}$, and $\tilde{\tau}_H^{\text{MC}}$,
where
\begin{eqnarray}
\label{eq:tilde_tau}
\tilde{\tau}_H^{\text{OW}} &=& \tau_H^\text{OW} \cdot (1-\tau_L), \quad
 \tilde{\tau}_H^{\text{MC}} = \tau_H^\text{MC} \cdot \tau_L .
\end{eqnarray}
In Eq.~(\ref{eq:tilde_tau}), $\tilde{\tau}_H^{\text{OW}}$ and $\tilde{\tau}_H^{\text{MC}}$ imply the probabilities that node H transmits in an OW period and in an MCOT period, respectively.
Since $\tau_H = \tilde{\tau}_H^\text{OW} + \tilde{\tau}_H^\text{MC}$,
we can compare $\tau_H$ with $\tilde{\tau}_H^{\text{OW}}$ and $\tilde{\tau}_H^{\text{MC}}$ in a fair basis.

\begin{figure}[!t]  
\begin{center}
\subfigure[$T_\text{MCOT}$ = 8 ms, RSF = 1]{
\includegraphics[width=0.466\columnwidth]{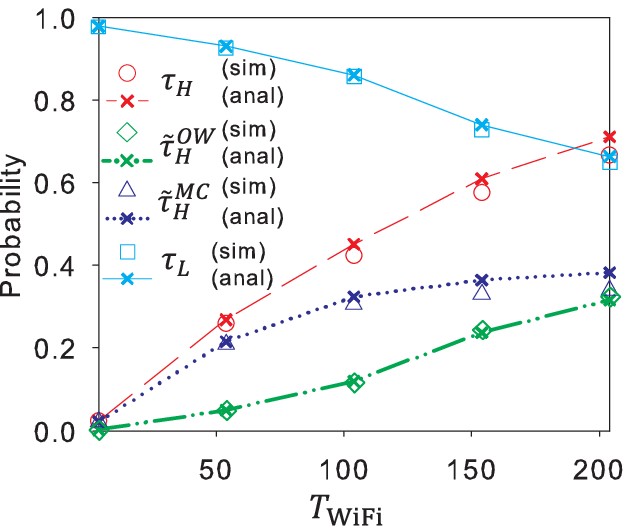}
\label{Fig:Transmit_Probability_8msec_RSF1}
}
\subfigure[$T_\text{MCOT}$ = 8 ms, RSF = 4]{
\includegraphics[width=0.466\columnwidth]{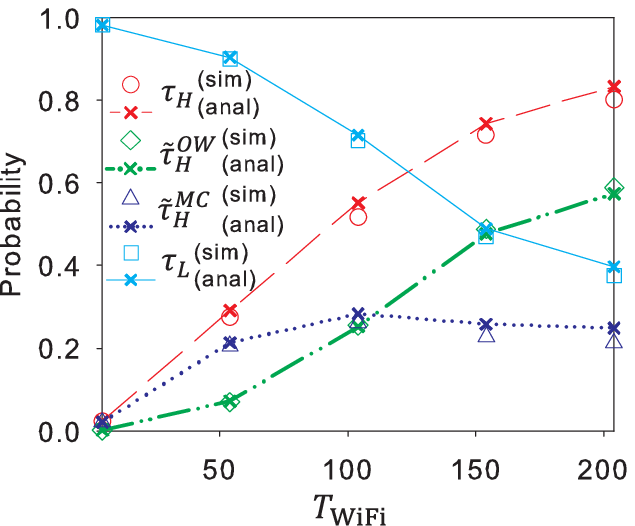}
\label{Fig:Transmit_Probability_8msec_RSF4}
}
\subfigure[$T_\text{MCOT}$ = 10 ms, RSF = 1]{
\includegraphics[width=0.466\columnwidth]{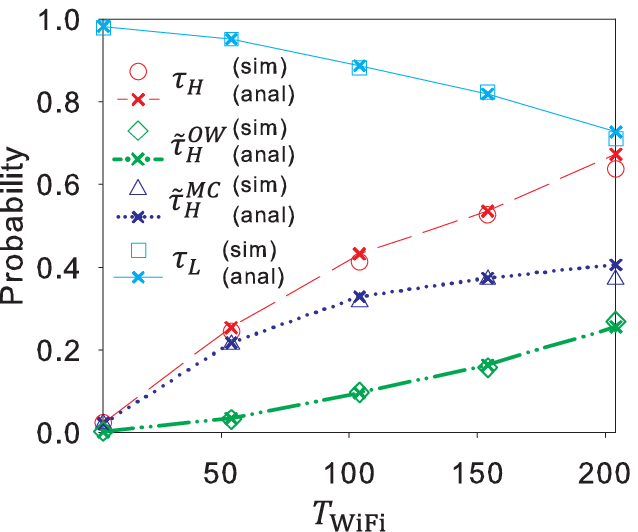}
\label{Fig:Transmit_Probability_10msec_RSF1}
}
\subfigure[$T_\text{MCOT}$ = 10 ms, RSF = 6]{
\includegraphics[width=0.466\columnwidth]{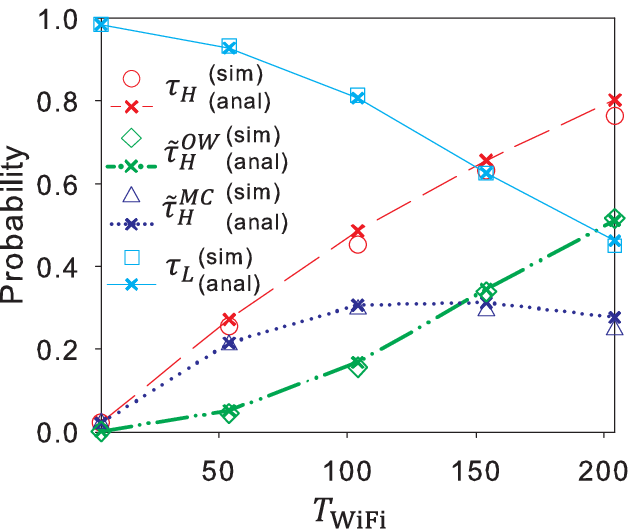}
\label{Fig:Transmit_Probability_10msec_RSF6}
}
\caption{Transmit probabilities: $\tau_L$, $\tau_H$, $\tilde{\tau}_H^{\text{OW}}$, and $\tilde{\tau}_H^{\text{MC}}$}
\label{Fig:Transmit_Probability}
\end{center}
\vspace{-0.15in}
\end{figure}

In Fig.~\ref{Fig:Transmit_Probability}, we tested two values of $T_\text{MCOT}$ and
two doubling policies (RSF 1 vs. RSF 4 or 6).
In every case, we can notice that the analytic results well match with the simulation, confirming the accuracy of the proposed model.
In addition, as $T_{\text{WiFi}}$ increases, $\tau_L$ keeps decreasing while $\tau_H$ keeps increasing.
That is, larger Wi-Fi packets tend to favor Wi-Fi's channel access while penalizing LTE-LAA, mainly due to the asymmetry between them.

Next, let us compare $\tau_L$ with $\tau_H$ in the following two cases:
\begin{itemize}
\item Case 1: when RSF is 1, in 
    Figs.~\ref{Fig:Transmit_Probability_8msec_RSF1} and \ref{Fig:Transmit_Probability_10msec_RSF1}, 
\item Case 2: when RSF is 4 or 6, in 
    Figs.~\ref{Fig:Transmit_Probability_8msec_RSF4} and \ref{Fig:Transmit_Probability_10msec_RSF6}.
\end{itemize}
As seen, case 1 is favorable for LTE-LAA such that node L almost always excels node H in terms of the transmit probability, thus impairing the fairness between the two technologies.
For fair coexistence, case 2 seems more reasonable since neither $\tau_L$ nor $\tau_H$ can outperform the other all the time.
In addition, we observe that case 2 makes $\tilde{\tau}_H^{\text{OW}}$ much enhanced but $\tilde{\tau}_H^{\text{MC}}$ suppressed, and sometimes it leads to $\tilde{\tau}_H^{\text{OW}} > \tilde{\tau}_H^{\text{MC}}$ at large $T_\text{WiFi}$.
This means that node H enjoys more successful transmission while experiencing less collision.

\begin{figure}[!t]
\begin{center}
\subfigure[$T_\text{MCOT}$ = 8 ms]{
\includegraphics[width=0.466\columnwidth]{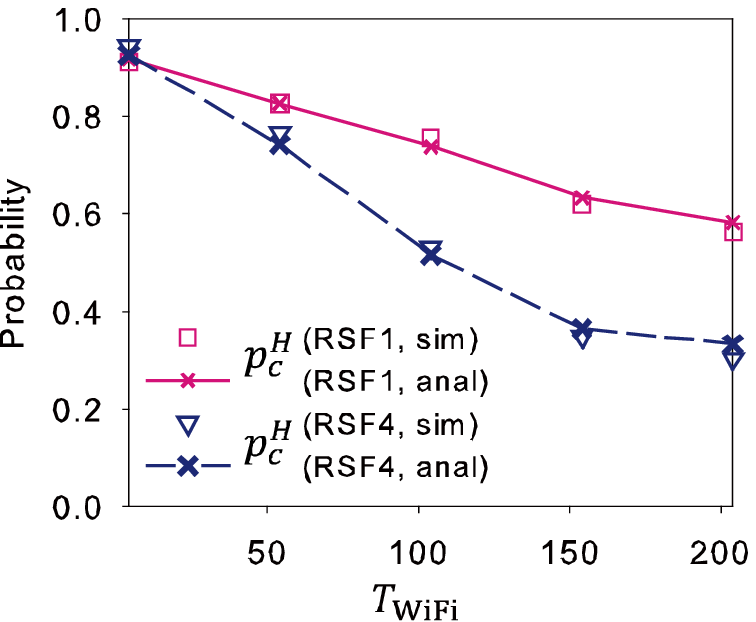}
\label{Fig:pcH_8msec}
}
\subfigure[$T_\text{MCOT}$ = 10 ms]{
\includegraphics[width=0.466\columnwidth]{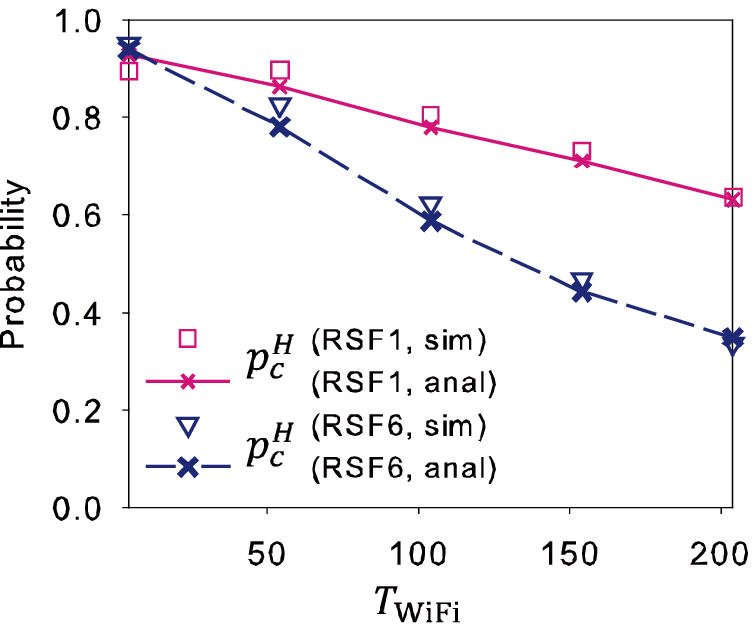}
\label{Fig:pcH_10msec}
}
\caption{Collision probability of node H: $p_c^H$}
\label{Fig:Collision_Pr}
\end{center}
\vspace{-0.15in}
\end{figure}

The difference between case 1 and case 2 arises due to the following reasons.
Within subframe 1, node H happens to transmit its packets only after the subframe starts, since an MCOT cannot begin in the middle of Wi-Fi's packet transmission.
A subframe other than 1, however, might overlap with a Wi-Fi packet that has started in the previous subframe.
Therefore, subframe 1 experiences less collision with node H's packets thus reducing $p_d$, and hence the doubling policy of RSF=1 allows node L to access the channel more aggressively.

The balance between $\tau_L$ and $\tau_H$ significantly varies with $T_\text{WiFi}$, and the value of $T_\text{WiFi}$ achieving $\tau_L = \tau_H$ changes with the chosen RSF.
Although smaller $T_\text{MCOT}$ tends to enhance the fairness, its impact seems less prominent than that of $T_\text{WiFi}$.
To promote fair coexistence, we can conclude that
the LTE-LAA standard needs to re-consider its current doubling policy with RSF=1, and
a proper RSF should be chosen according to the average packet duration of Wi-Fi.

\subsection{Collision and Doubling Probabilities}
\label{subsec:cbdPr}
\begin{figure}[!t]
\begin{center}
\subfigure[$T_\text{MCOT}$ = 8 ms]{
\includegraphics[width=0.466\columnwidth]{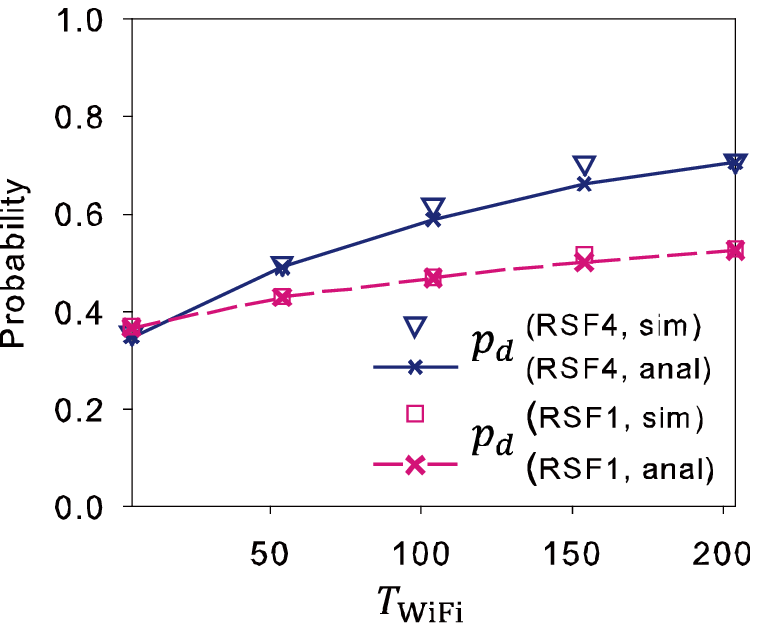}
\label{Fig:BusyDoublingPr_8msec}
}
\subfigure[$T_\text{MCOT}$ = 10 ms]{
\includegraphics[width=0.466\columnwidth]{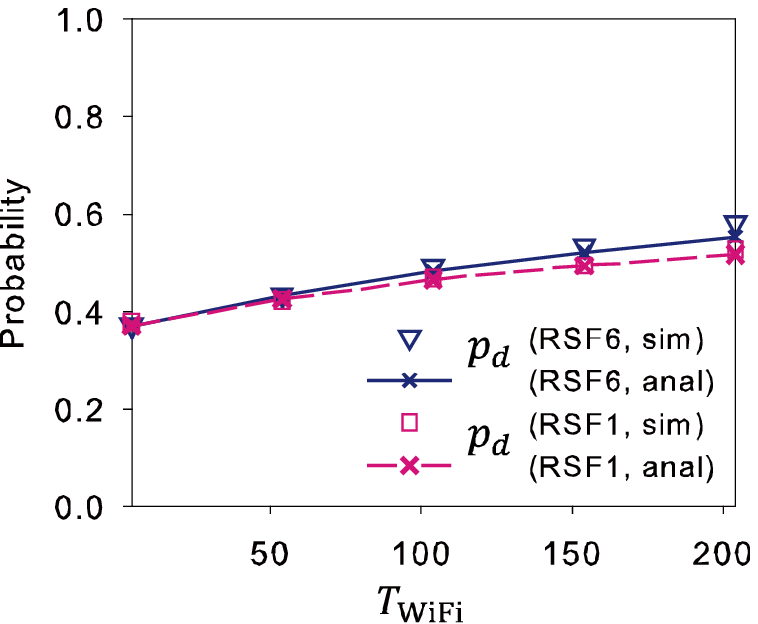}
\label{Fig:BusyDoublingPr_10msec}
}
\caption{Doubling probability of node L: $p_d$}
\label{Fig:DoublingPr}
\end{center}
\vspace{-0.1in}
\end{figure}

Fig.~\ref{Fig:Collision_Pr} shows the collision probability of node H, $p_c^H$, and
Fig.~\ref{Fig:DoublingPr} presents the doubling probability of node L, $p_d$.
The plots confirm that the analysis well matches with the simulation results.
%
Table~\ref{table:AvgCorruptedSF} also investigates the accuracy of the model in terms of the average number of subframes in collision per MCOT, where the error is well bounded within the range from -5\% to -1\%.

\begin{table}[!t]
\caption{Average number of subframes in collision per MCOT}
\vspace{-0.1in}
\label{table:AvgCorruptedSF}
\begin{center}
\begin{tabularx}{1\columnwidth}{|c|>{\centering}X|>{\centering}X|>{\centering}X|X|X|}
\cline{1-5}
\multirow{2}{*}{$T_\text{WiFi}$} & \multicolumn{2}{c|}{$T_\text{MCOT} =$ 8 ms, RSF=1}& \multicolumn{2}{c|}{$T_\text{MCOT} =$ 8 ms, RSF=4}\\ \cline{2-5}
			&Simulation		&Analysis		&Simulation		& \centering Analysis \tabularnewline
\cline{1-5}
4 			&  2.83   			&  2.83		&  2.83   			&  \centering 2.82 \tabularnewline
54 			&  3.86			&  3.84		&  3.90			&  \centering 3.86 \tabularnewline
104 			&  4.57			&  4.48		&  4.72			&  \centering 4.57 \tabularnewline
154 			&  5.08			&  4.93		&  5.32			&  \centering 5.08 \tabularnewline
204 			&  5.43			&  5.24		&  5.68			&  \centering 5.41 \tabularnewline
\cline{1-5}
\end{tabularx}
\end{center}
\vspace{-0.1in}
\end{table}

\begin{figure*}[!t]
\begin{center}
\subfigure[$T_\text{WiFi}=4$]{
\includegraphics[width=0.19\textwidth]{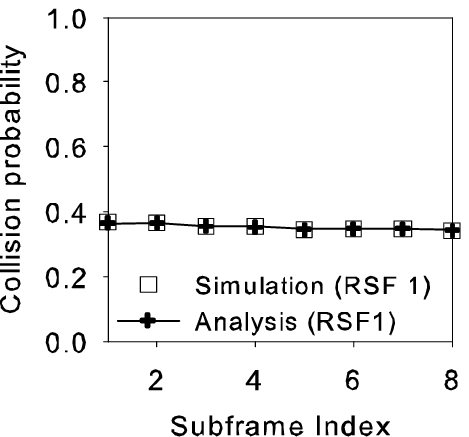}
\label{fig:SF_Collision_RSF1_TxTime_4}
}
\subfigure[$T_\text{WiFi}=54$]{
\includegraphics[width=0.1725\textwidth]{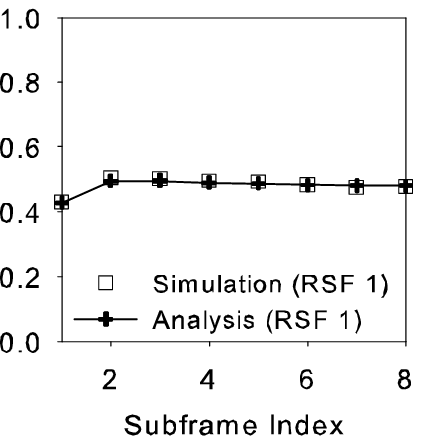}
\label{fig:SF_Collision_RSF1_TxTime_54}
}
\subfigure[$T_\text{WiFi}=104$]{
\includegraphics[width=0.1725\textwidth]{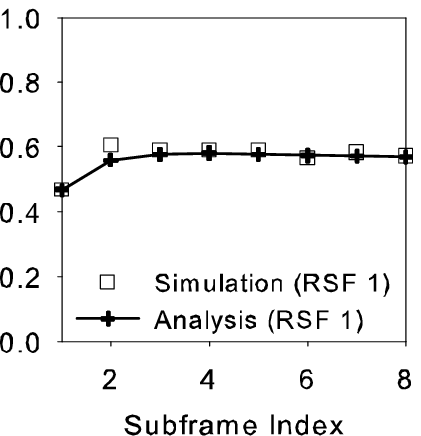}
\label{fig:SF_Collision_RSF1_TxTime_104}
}
\subfigure[$T_\text{WiFi}=154$]{
\includegraphics[width=0.1725\textwidth]{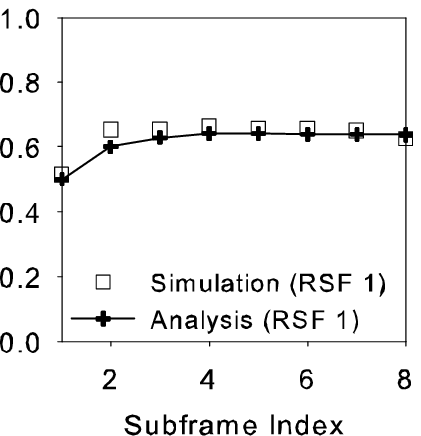}
\label{fig:SF_Collision_RSF1_TxTime_154}
}
\subfigure[$T_\text{WiFi}=204$]{
\includegraphics[width=0.1725\textwidth]{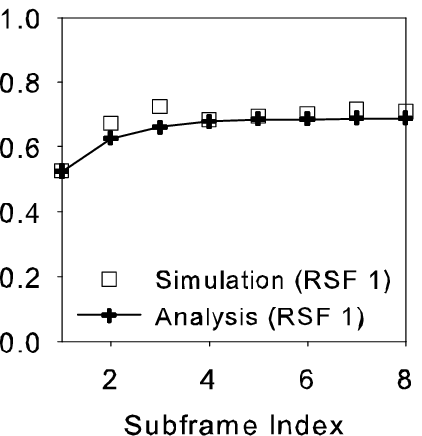}
\label{fig:SF_Collision_RSF1_TxTime_204}
}
\caption{Subframe collision probability (when $T_\text{MCOT} = 8$ ms, RSF = 1)}
\label{Fig:SF_Collision_8msec_RSF_1}
\end{center}
\vspace{-0.1in}
\end{figure*}

\begin{figure*}[!t]
\begin{center}
\subfigure[$T_\text{WiFi}=4$]{
\includegraphics[width=0.19\textwidth]{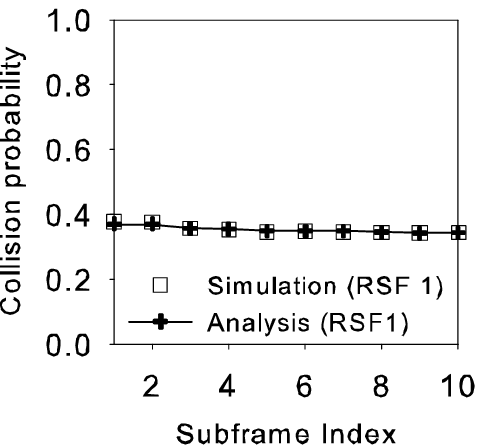}
\label{fig:SF_Collision_RSF6_TxTime_4}
}
\subfigure[$T_\text{WiFi}=54$]{
\includegraphics[width=0.1725\textwidth]{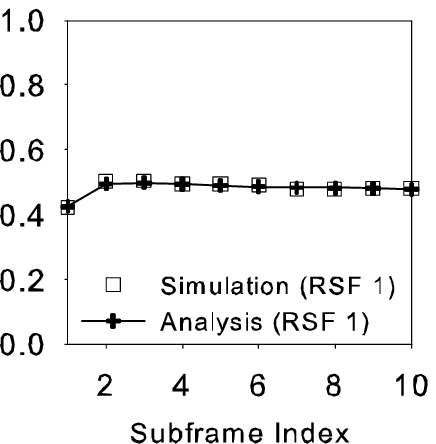}
\label{fig:SF_Collision_RSF6_TxTime_54}
}
\subfigure[$T_\text{WiFi}=104$]{
\includegraphics[width=0.1725\textwidth]{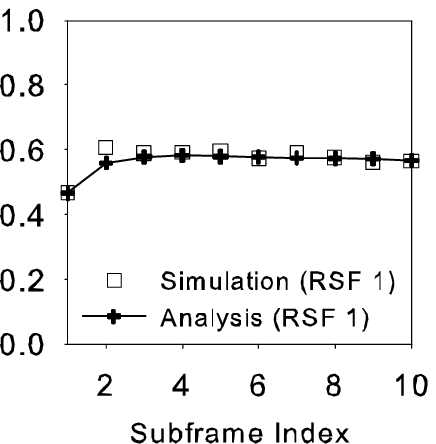}
\label{fig:SF_Collision_RSF6_TxTime_104}
}
\subfigure[$T_\text{WiFi}=154$]{
\includegraphics[width=0.1725\textwidth]{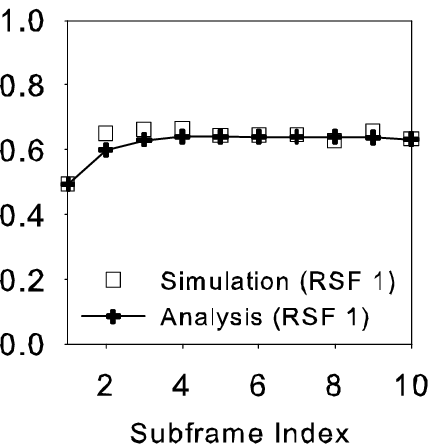}
\label{fig:SF_Collision_RSF6_TxTime_154}
}
\subfigure[$T_\text{WiFi}=204$]{
\includegraphics[width=0.1725\textwidth]{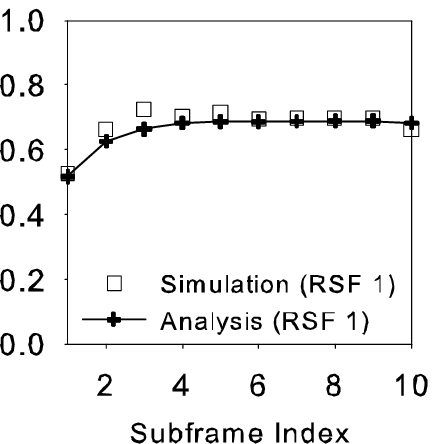}
\label{fig:SF_Collision_RSF6_TxTime_204}
}
\caption{Subframe collision probability (when $T_\text{MCOT} = 10$ ms, RSF = 1)}
\label{Fig:SF_Collision_10msec_RSF_1}
\end{center}
\vspace{-0.1in}
\end{figure*}

It is observed that as $T_\text{WiFi}$ increases, $p_c^H$ keeps decreasing while $p_d$ keeps increasing.
In other words, as node H consumes longer airtime per transmission,
a Wi-Fi packet more easily collides with a subframe (i.e., larger $p_d$).
Table~\ref{table:AvgCorruptedSF} also supports this by showing an increasing trend in the number of collided subframes with $T_\text{WiFi}$.
Then, increased $p_d$ results in more reluctant transmission by node L, and in turn node H will encounter MCOTs less frequently thus leading to enhanced $p_c^H$.

In Fig.~\ref{Fig:Collision_Pr}, changing RSF from 1 to the last eligible one (4 or 6) significantly improves $p_c^H$.
Such a change, however, penalizes node L by increasing $p_d$ moderately, due to the reason discussed in Section~\ref{subsec:TxPr}.
The impact of setting RSF on $p_c^H$ can be explained as follows.
Increasing $p_d$ implies more reluctant transmission by node L, and thus node H encounters MCOT periods less frequently.
Since node H experiences collision only during an MCOT, $p_c^H$ gets smaller.

On the other hand, for any given RSF, reducing $T_\text{MCOT}$ slightly decreases $p_c^H$, thus promoting more successful transmission of Wi-Fi.
As a result, there is a tradeoff between the collision probability and the doubling probability, which should be also considered in determining proper $T_\text{MCOT}$ and RSF.
Among $T_\text{MCOT}$ and RSF, however, RSF turns out to be more influential.

\begin{table}[!t]
\caption{The value of $z_2$}
\vspace{-0.1in}
\label{table:z2}
\begin{tabularx}{1\columnwidth}{|c|>{\centering}X|>{\centering}X|>{\centering}X|X|X|}
\cline{1-5}
\multirow{2}{*}{$T_\text{WiFi}$} & \multicolumn{2}{c|}{$T_\text{MCOT}$ = 8 ms} & \multicolumn{2}{c|}{$T_\text{MCOT}$ = 10 ms} \\ \cline{2-5}
  	 	         & RSF=1       & RSF=4             & RSF=1        &  \centering RSF=6      \tabularnewline\cline{1-5}
4             		     & 0.496         & 0.493              & 0.503         &  \centering 0.494       \tabularnewline
54              	     & 0.500         & 0.498              & 0.499         &  \centering 0.501       \tabularnewline
104           		     & 0.497         & 0.495              & 0.501          &  \centering 0.502       \tabularnewline
154                        & 0.507         & 0.509               & 0.496         &  \centering 0.492       \tabularnewline
204             	     & 0.436         & 0.412               & 0.507        &  \centering 0.519       \tabularnewline
\cline{1-5}
\end{tabularx}
\vspace{-0.1in}
\end{table}
When we derived $p_c^H$ in Section~\ref{sec:NodeHcollision}, $z_1 = z_2 = 1/2$ was assumed.
Table~\ref{table:z2} lists the value of $z_2$ measured by simulation, from which we notice that the assumption is quite reasonable in most cases except when $T_\text{WiFi} = 204$ and $T_\text{MCOT} = 8$ ms.
This is because when $T_\text{WiFi}$ gets more comparable to $T_\text{MCOT}$, e.g., 204 slots versus 888 slots in the 8 ms case, the moment the overlap event happens within $T_\text{WiFi}$ starts to be correlated.
One straightforward (but not realizable) example is $T_\text{WiFi} \gg T_\text{MCOT}$, where $z_2 \gg z_1$.
In practice, however, $T_\text{WiFi}$ is upperbounded so that we can still assume
$z_1 = z_2 = 1/2$.

\subsection{Subframe Collision Probabilities}
\label{subsec:Csf}

Figs.~\ref{Fig:SF_Collision_8msec_RSF_1} and \ref{Fig:SF_Collision_10msec_RSF_1} present the subframe collision probability $C_{\text{sf}}$ with various $T_\text{WiFi}$ and two $T_\text{MCOT}$ values.
%
%
It can be seen that our analysis well matches the simulation results, where the error between them tends to increase with $T_\text{WiFi}$.
%
Such a gap is rooted at the fact that node H's MC in Fig.~\ref{Fig:MC_WiFi} assumes that $p_c^H$ is constant and independent of node H's backoff stage.
In fact, $p_c^H$ may depend on the backoff stage due to the following reasons.
Since node H's transmissions in OW periods are mostly successful (except in the case of the overlap event), node H's backoff stage stays at the smallest stage with high probability.
On the contrary, during an MCOT, node H experiences frequent collisions and thus its backoff stage keeps increasing while sometimes resetting to the minimum whenever reaching the maximum retry limit.
Therefore, the minimum backoff stage would likely be involved with many successful transmissions in OW periods and a few collisions within MCOTs, resulting in relatively smaller per-stage collision probability.
Although the error in the $C_{\text{sf}}$ analysis seems well bounded, one can try to enhance $p_c^H$ to become stage-dependent for better accuracy, which is left as our future work.

Fixing the subframe index, $C_{\text{sf}}$ tends to increase with $T_\text{WiFi}$ and eventually becomes fairly large at $T_\text{WiFi} = 204$.
This is a natural phenomenon since a longer Wi-Fi packet overlaps with more subframes.
While fixing $T_\text{WiFi}$, however, the evolution of $C_{\text{sf}}(r)$ with varying $r$ appears differently according to which value of $T_\text{WiFi}$ is given.
When $T_\text{WiFi} = 4$, $C_{\text{sf}}$ keeps decreasing.
For other $T_\text{WiFi}$, $C_{\text{sf}}(r)$ increases until $r$ reaches 2 or 3, and then starts to decrease afterwards.

%

The subframe-dependent pattern of $C_{\text{sf}}$ suggests that LTE-LAA should carefully decide which subframe a downlink traffic is assigned, since each subframe offers a different chance of collision by Wi-Fi.
Without asymmetry, such consideration would not be necessary because collision between LTE and Wi-Fi occurs only when both Wi-Fi's packet transmission and an MCOT period start simultaneously, where the impact of collision is confined to the first subframe (and possibly in its following subframes in case $T_\text{WiFi}$ is fairly large).
%
In asymmetric coexistence, however, node H can transmit in any subframe since it cannot see LTE, where the subframe in which node H transmits is decided by its backoff stage evolution and $T_\text{WiFi}$.

\subsection{Throughput and Channel Access Delay}
\label{subsec:Throughput_and_Delay}
\begin{figure}[!t]
\begin{center}
\subfigure[$T_\text{MCOT}$ = 8 ms, RSF = 1]{
\includegraphics[width=0.466\columnwidth]{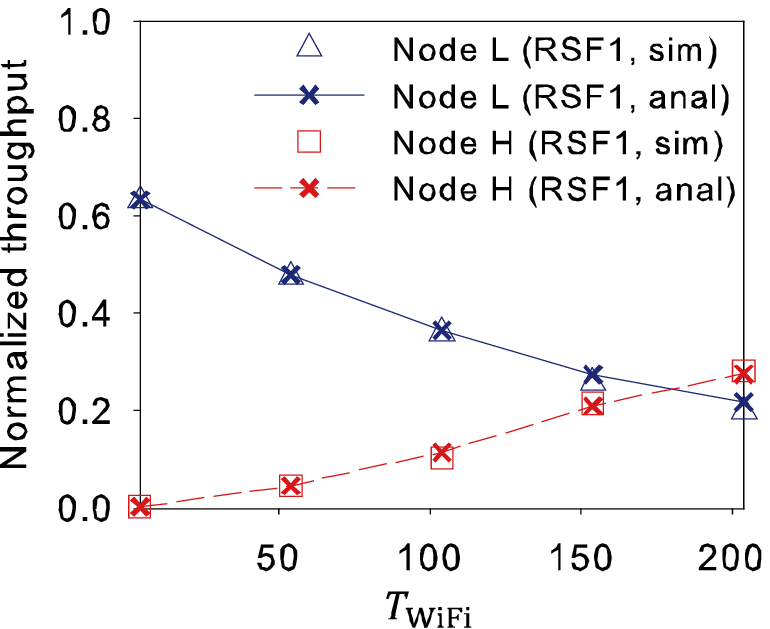}
\label{Fig:Throughput_8msec_RSF1}
}
\subfigure[$T_\text{MCOT}$ = 8 ms, RSF = 4]{
\includegraphics[width=0.466\columnwidth]{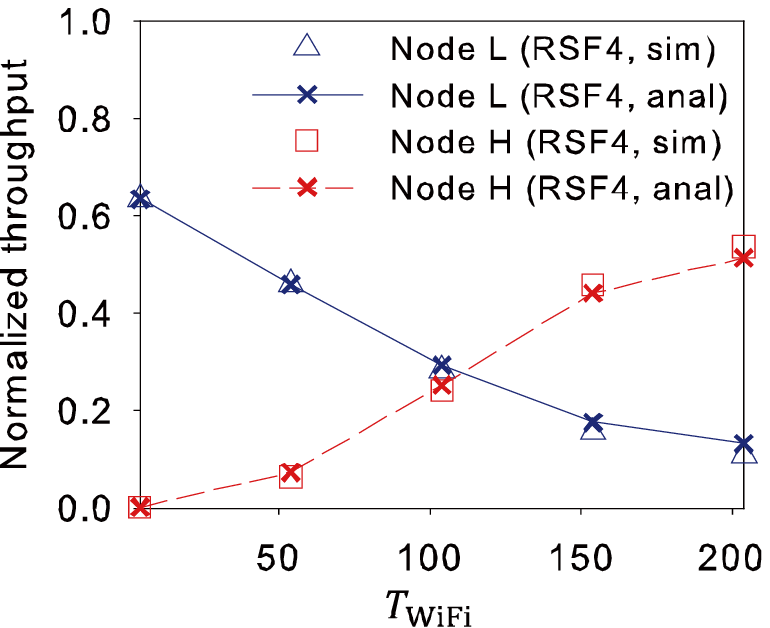}
\label{Fig:Throughput_8msec_RSF4}
}
\subfigure[$T_\text{MCOT}$ = 10 ms, RSF = 1]{
\includegraphics[width=0.466\columnwidth]{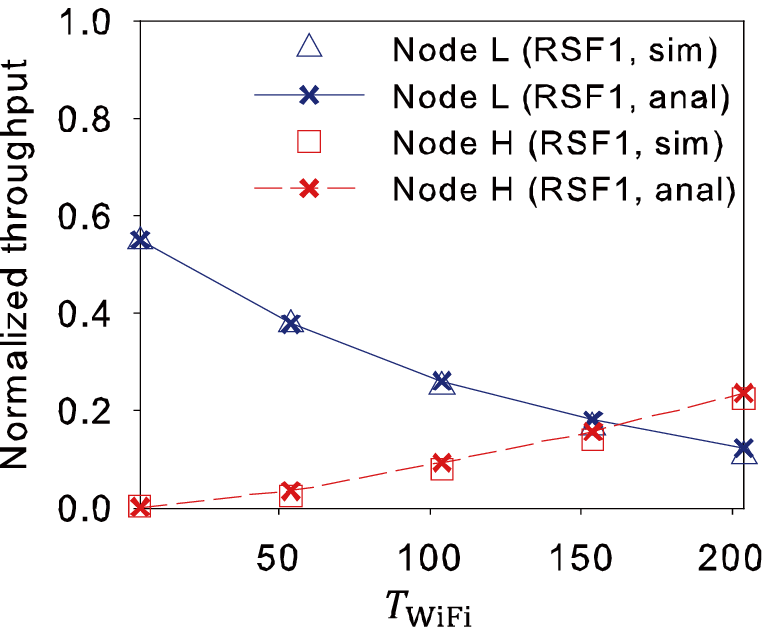}
\label{Fig:Throughput_10msec_RSF1}
}
\subfigure[$T_\text{MCOT}$ = 10 ms, RSF = 6]{
\includegraphics[width=0.466\columnwidth]{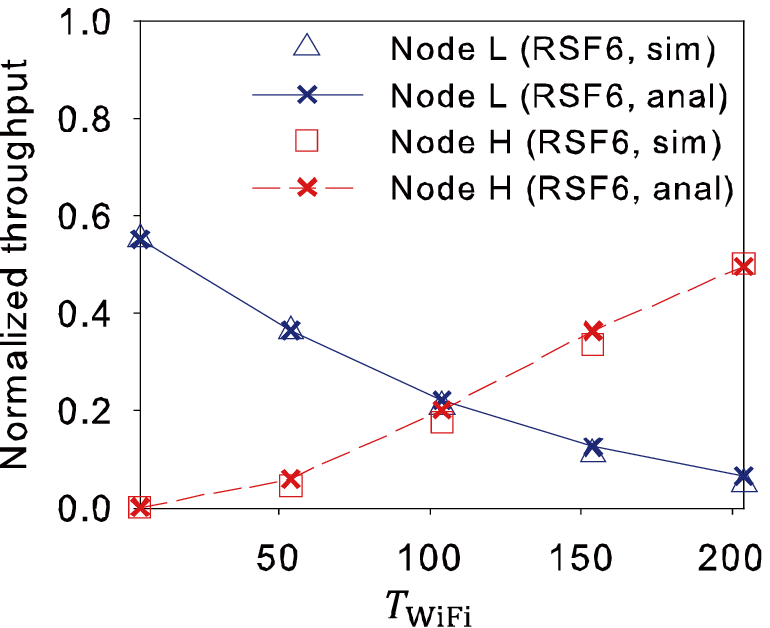}
\label{Fig:Throughput_10msec_RSF6}
}
\caption{Throughput of node L and node H ($S_L$ and $S_H$)}
\label{Fig:Throughput}
\end{center}
\vspace{-0.1in}
\end{figure}

\begin{figure}[!t]
\begin{center}
\subfigure[$T_\text{MCOT}$ = 8 ms, RSF = 1]{
\includegraphics[width=0.466\columnwidth]{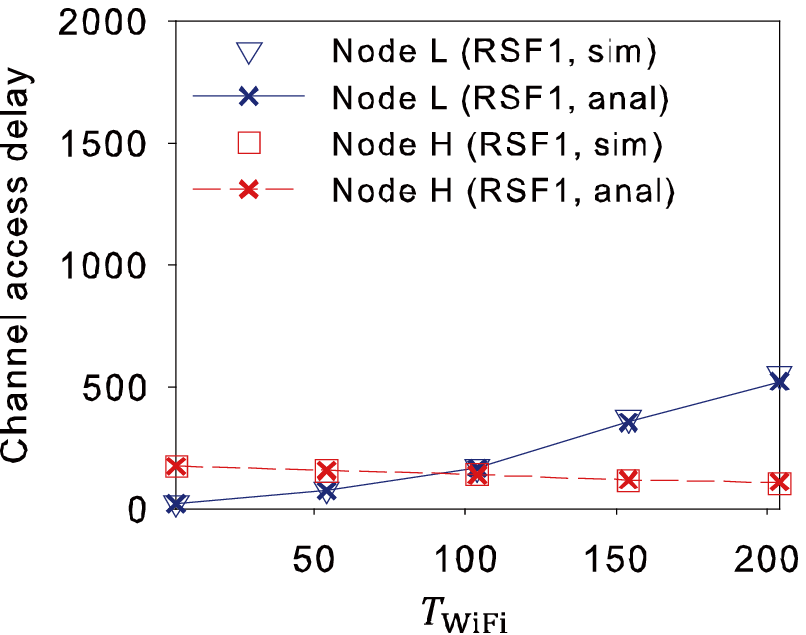}
\label{Fig:Delay_8msec_RSF1}
}
\subfigure[$T_\text{MCOT}$ = 8 ms, RSF = 4]{
\includegraphics[width=0.466\columnwidth]{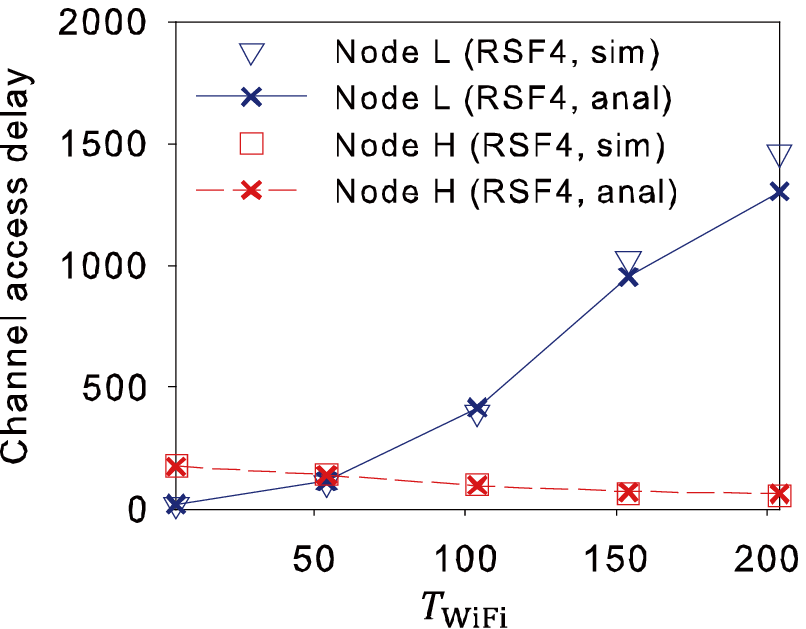}
\label{Fig:Delay_8msec_RSF4}
}
\subfigure[$T_\text{MCOT}$ = 10 ms, RSF = 1]{
\includegraphics[width=0.466\columnwidth]{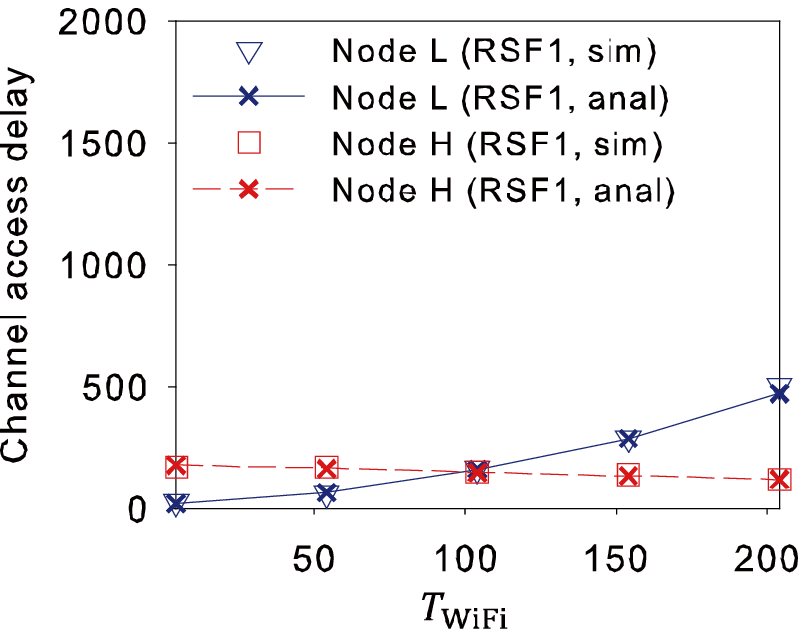}
\label{Fig:Delay_10msec_RSF1}
}
\subfigure[$T_\text{MCOT}$ = 10 ms, RSF = 6]{
\includegraphics[width=0.466\columnwidth]{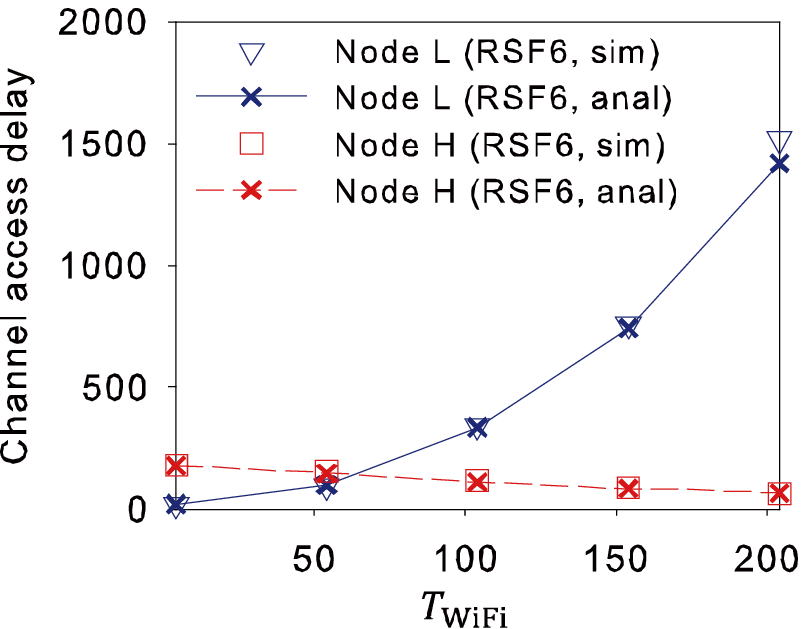}
\label{Fig:Delay_10msec_RSF6}
}
\caption{Average channel access delay of node L and node H ($E[D_L]$ and $E[D_H]$)}
\label{Fig:Delay}
\end{center}
\end{figure}

Figs.~\ref{Fig:Throughput} and \ref{Fig:Delay} present the normalized throughput ($S_L$ and $S_H$) and the average channel access delay ($E[D_L]$ and $E[D_H]$) of node L and node H, with various combinations of $T_\text{MCOT}$ and RSF.
All results in the figures confirm the close match between the analysis and the simulation.
In Fig.~\ref{Fig:Throughput}, as $T_\text{WiFi}$ grows, $S_L$ keeps degraded while $S_H$ is consistently enhanced, meaning that longer Wi-Fi packets suppress LTE-LAA's throughput.
In the same vein, Fig.~\ref{Fig:Delay} shows a similar but reversed behavior regarding the average channel access delay with varying $T_{\text{WiFi}}$.
Notably, however, the impact on the delay seems much more prominent for LTE-LAA than Wi-Fi because node L should always wait for the completion of node H's ongoing transmission before resuming its backoff procedure.

In Figs.~\ref{Fig:Throughput_8msec_RSF1} and \ref{Fig:Throughput_10msec_RSF1}, RSF$=1$ is favorable for node L, such that node L almost always excels node H in terms of the throughput.
For fair coexistence, RSF of 4 and 6 seem more reasonable since neither $S_L$ nor $S_H$ can outperforms the other.
The difference between RSF$=1$ and RSF of 4 or 6 arises due to the same reasons causing the difference between case 1 and case 2 in Section~\ref{subsec:TxPr}.
Similar to Section~\ref{subsec:TxPr}, the ratio of $S_L$ to $S_H$ fluctuates with $T_\text{WiFi}$, and the value of $T_\text{WiFi}$ that makes the two throughput even is very sensitive to the RSF while less so to $T_\text{MCOT}$.
To achieve fairness in throughput, we can conclude that RSF$=1$ is not the best choice, and the RSF should be adaptively chosen according to $T_\text{WiFi}$.

In Figs.~\ref{Fig:Delay_8msec_RSF1} and \ref{Fig:Delay_8msec_RSF4},
the gap between the minimum and the maximum values of $E[D_L]$ becomes significantly increased when we change the RSF from 1 to 4, causing excessively longer channel access delay for node L.
The same phenomenon is observed in Figs.~\ref{Fig:Delay_10msec_RSF1} and \ref{Fig:Delay_10msec_RSF6} between RSF of 1 and 6.
As shown in Section~\ref{subsec:Csf}, later subframes like 4 and 6 experience more collision caused by node H's transmission, and hence the doubling policy of RSF$=4$ or $6$ makes node L undergo longer channel access delay due to increased $p_d$.
Contrary to the case of throughput, RSF of 1 can be a better solution for fair coexistence in terms of delay.

\begin{figure}[!t]
\begin{center}
\includegraphics[width=0.7\columnwidth]{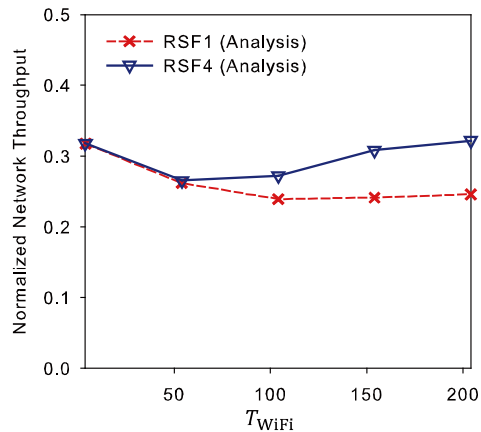}
\caption{Sum throughput $S_L + S_H$}
\label{Fig:Network_Throughput}
\end{center}
\vspace{-0.1in}
\end{figure}

In Figs.~\ref{Fig:Delay_8msec_RSF4} and \ref{Fig:Delay_10msec_RSF6}, $E[D_L]$ gets improved (i.e., reduced) by extending $T_\text{MCOT}$ from 8 ms to 10 ms.
Although somewhat subtle, the same is observed from Figs.~\ref{Fig:Delay_8msec_RSF1} and \ref{Fig:Delay_10msec_RSF1}.
The difference between 8 ms and 10 ms is caused by the following reason.
In Section~\ref{subsec:TxPr}, we observed that $\tau_L$ decreases as $T_\text{MCOT}$ becomes smaller.
The decrement of $\tau_L$ results in sparser access of channel, which results in longer delay.
Deduced from the above discussion, we can conclude that larger $T_\text{MCOT}$ is better for achieving more balanced delay between LTE-LAA and Wi-Fi.

In summary, there exists a tradeoff between throughput and delay in achieving fair coexistence, where the RSF and $T_\text{MCOT}$ are the two control knobs.
For the perspective of throughput, the doubling policy should prefer the last eligible subframe (4 or 6) as the RSF, while pursuing smaller $T_\text{MCOT}$ is also favorable but with a limited impact.
In addition, Fig.~\ref{Fig:Network_Throughput} shows that RSF 4 also achieves larger sum throughput (i.e., $S_L + S_H$) than RSF 1.
From the viewpoint of delay, however, having the RSF as subframe 1 and larger $T_\text{MCOT}$ are certainly desirable, where the RSF is the more dominant factor.
Therefore, a network operator of LTE-LAA should carefully determine the doubling policy (or equivalently, the RSF) and $T_\text{MCOT}$, considering the QoS requirement of the given traffic, e.g., capacity craving vs. delay sensitive.



\section{Discussion}
\label{sec:discussion}

This section discusses additional issues LTE-LAA and Wi-Fi should consider in modeling and designing their systems.


Wi-Fi adopts rate adaptation algorithms like ARF \cite{Kamerman97}, AARF \cite{Lacage04}, and SampleRate \cite{Bicket05}.
These algorithms decrease the data rate when consecutive collisions occur, and increase the data rate when a certain condition is met.
The condition for increment, however, is designed stricter than that for decrement.
Hence, when node H experiences multiple collisions during an MCOT period,
it could stay with a low data rate. 
Since a low data rate implies longer per-packet airtime, Wi-Fi with rate adaptation may incur longer delay to LTE-LAA.


Recent Wi-Fi standards support packet aggregation as found in IEEE 802.11n \cite{80211n}.
Packet aggregation merges multiple packets into one, and hence it has an effect to make the airtime of Wi-Fi longer.
With aggressive packet aggregation, Wi-Fi will eventually occupy the channel more than LTE-LAA thus impairing the fairness.

The above discussions suggest that the impact of rate adaptation and packet aggregation should be considered when LTE-LAA designs its operational parameters.
Although such considerations are left as our future work, the proposed model in this paper can indirectly address this by setting $T_\text{WiFi}$ as the average airtime of Wi-Fi.

\section{Conclusion and Future Work}
\label{sec:conclusion}
In this work, we captured the asymmetric hidden terminal problem in an LAA network by modeling the network with a joint MC, and derived its stationary probabilities along with the key performance metrics.
Via extensive numerical evaluations, we have shown the accuracy of the proposed model and the impact of the asymmetric hidden terminal.
In the future, we would like to extend the model to a more general setup, where multiple exposed and hidden Wi-Fi STAs co-exist with the LTE-LAA eNB.
In addition, we want to enhance our model to consider the slight difference in DIFS between LTE-LAA and Wi-Fi \cite{3GPPTS36.213}.


\section*{Acknowledgments}
This work was supported by Institute for Information \& communications Technology Promotion(IITP) grant funded by the Korea government(MSIP) (No. R0101-15-244, Development of 5G Mobile Communication Technologies for Hyper-connected smart services).

\bibliographystyle{ieeetr}
\bibliography{references}

\begin{thebibliography}{10}

\bibitem{Cisco16}
Cisco, ``{Cisco Visual Networking Index: Global Mobile Data Traffic Forecast
  Update, 2015–2020},'' February 2016.

\bibitem{3GPPTS23.207}
3GPP, ``{End-to-end Quality of Service (QoS) concept and architecture},''
  December 2009.
\newblock TS 23.207 V9.0.0.

\bibitem{3GPPTR36.889}
3GPP, ``{3rd Generation Partnership Project; Technical Specification Group
  Radio Access Network; Study on Licensed-Assisted Access to Unlicensed
  Spectrum; (Release 13)},'' June 2015.
\newblock TR 36.889 V13.0.0.

\bibitem{80211}
``{Part 11: Wireless LAN Medium Access Control (MAC) and Physical Layer (PHY)
  specifications},'' March 2012.

\bibitem{Ekici08}
O.~Ekici and A.~Yongacoglu, ``{IEEE 802.11a Throughput Performance with Hidden
  Nodes},'' {\em IEEE Communications Letters}, vol.~12, pp.~465--467, June
  2008.

\bibitem{Hung10}
F.~Hung and I.~Marsic, ``{Performance analysis of the IEEE 802.11 DCF in the
  presence of the hidden stations},'' {\em Computer Networks}, vol.~54,
  pp.~2674--2687, October 2010.

\bibitem{Tsertou08}
A.~Tsertou and D.~Laurenson, ``{Revisiting the hidden terminal problem in a
  CSMA/CA wireless network},'' {\em {IEEE} Trans. Mobile Comput.}, vol.~7,
  pp.~817 -- 831, July 2008.

\bibitem{Jeon14}
J.~Jeon, Q.~Li, H.~Niu, A.~Papathanassiou, and G.~Wu, ``{LTE in the Unlicensed
  Spectrum: A Novel Coexistence Analysis with WLAN Systems},'' in {\em Proc.
  {IEEE} {Globecom}}, 2014.

\bibitem{Bhorkar14}
A.~Bhorkar, C.~Ibars, and P.~Zong, ``{Performance Analysis of LTE and Wi-Fi in
  Unlicensed Band Using Stochastic Geometry},'' in {\em Proc. {IEEE} {PIMRC}},
  2014.

\bibitem{Mvulla15}
J.~Mvulla, E.~Park, M.~Adnan, and J.~Son, ``{Analysis of Asymmetric Hidden Node
  Problem in IEEE 802.11ax Heterogeneous WLANs},'' in {\em Proc. {IEEE}
  {ICTC}}, 2015.

\bibitem{Bianchi00}
G.~Bianchi, ``{Performance Analysis of the IEEE 802.11 Distributed Coordination
  Function},'' {\em J. Sel. Areas Commun.}, vol.~18, pp.~535--547, March 2000.

\bibitem{Ziouva02}
E.~Ziouva and T.~Antonakopoulos, ``{CSMA/CA performance under high traffic
  conditions: throughput and delay analysis},'' {\em Computer Communications},
  vol.~25, pp.~313--321, February 2002.

\bibitem{Malone07}
D.~Malone, K.~Duffy, and D.~Leith, ``{Modeling the 802.11 distributed
  coordination function in nonsaturated heterogeneous conditions},'' {\em
  {IEEE}/{ACM} Trans. Networking}, vol.~15, pp.~159--172, February 2007.

\bibitem{Song16}
Y.~Song, K.~Sung, and Y.~Han, ``{Coexistence of Wi-Fi and Cellular With
  Listen-Before-Talk in Unlicensed Spectrum},'' {\em IEEE Communications
  Letters}, vol.~20, pp.~161--164, January 2016.

\bibitem{Chen15}
C.~Chen, R.~Ratasuk, and A.~Ghosh, ``{Downlink Performance Analysis of LTE and
  WiFi Coexistence in Unlicensed Bands with a Simple Listen-before-talk
  Scheme},'' in {\em Proc. {IEEE} {VTC Spring}}, 2015.

\bibitem{Cano16}
C.~Cano and D.~Leith, ``{Unlicensed LTE/WiFi Coexistence: Is LBT Inherently
  Fairer Than CSAT?},'' in {\em Proc. {IEEE} {ICC}}, 2016.

\bibitem{Zhang15}
R.~Zhang, M.~Wang, L.~Cai, X.~Shen, L.~Xie, and Y.~Cheng, ``{Modeling and
  Analysis of MAC Protocol for LTE-U Co-existing with Wi-Fi},'' in {\em Proc.
  {IEEE} {Globecom}}, 2015.

\bibitem{Yin15}
R.~Yin, G.~Yu, A.~Maaref, and G.~Li, ``{Adaptive LBT for Licensed Assisted
  Access LTE Networks},'' in {\em Proc. {IEEE} {Globecom}}, 2015.

\bibitem{3GPP-RP-151569}
3GPP, ``{Release 13 analytical view version},'' September 2015.
\newblock RP-151569.

\bibitem{ITU-Report15}
ITU, ``{IMT traffic estimates for the years 2020 to 2030},'' July 2015.

\bibitem{LTE-U-Forum}
L.-U. Forum, ``{LTE-U Technical Report - Coexistence Study for LTE-U SDL},''
  February 2015.

\bibitem{LTE_Small_Book}
J.~R. Harri~Holma, Antti~Toskala, ``{LTE Small Cell Optimization: 3GPP
  Evolution to Release 13},'' November 2015.

\bibitem{ETSI301893}
ETSI, ``{Broadband Radio Access Networks (BRAN); 5 GHz high performance RLAN;
  Harmonized EN covering the essential requirements of article 3.2 of the
  R\&TTE Directive},'' March 2015.

\bibitem{Babich10}
F.~Babich and M.~Comisso, ``{Theoretical Analysis of Asynchronous Multi-Packet
  Reception in 802.11 Networks},'' {\em IEEE Trans. Commun.}, vol.~58,
  pp.~1782--1794, June 2010.

\bibitem{3GPPTS36.213}
3GPP, ``{3rd Generation Partnership Project; Technical Specification Group
  Radio Access Network; Evolved Universal Terrestrial Radio Access (E-UTRA);
  Physical layer procedures (Release 13)},'' June 2016.
\newblock TS 36.213 V13.2.0.

\bibitem{Bai11}
J.~Bai and P.~Wang, ``{Conditional Markov chain and its application in economic
  time series analysis},'' {\em Journal of Applied Econometrics}, vol.~26,
  pp.~715--734, August 2011.

\bibitem{80211ac}
{IEEE 802.11ac-2013}, ``{Part 11: Wireless LAN Medium Access Control (MAC) and
  Physical Layer (PHY) specifications; Amendment 4: Enhancements for Very High
  Throughput for Operation in Bands below 6 GHz},'' December 2013.

\bibitem{Kamerman97}
A.~Kamerman and L.~Monteban, ``{WaveLAN-II: A high-performance wireless lan for
  the unlicensed band},'' in {\em AT\&T Bell Laboratories Technical Journal},
  pp.~118--133, 1997.

\bibitem{Lacage04}
H.~M. M.~Lacage and T.~Turletti, ``{IEEE 802.11 rate adaptation: A practical
  approach},'' in {\em Institut National De Recherche en Informatique et en
  Auomatique}, 2004.

\bibitem{Bicket05}
J.~Bicket, ``{ Bit-rate Selection in Wireless Networks},'' in {\em MIT Master's
  Thesis}, 2005.

\bibitem{80211n}
{IEEE 802.11n-2009}, ``{Part 11: Wireless LAN Medium Access Control (MAC)and
  Physical Layer (PHY) Specifications Amendment 5: Enhancements for Higher
  Throughput},'' October 2009.

\end{thebibliography}

\end{document}